# Smooth-Curvature Bend Design Guided by Variational Analysis for Adiabatic Multimode Integrated Photonics

**Yutong Shi[1,2,*], Shijin Yu[1,2,*], Ye Huang[1,2], Yuan Yu[1,2,3,✉], and Xinliang Zhang[1,2,3,✉]**

[1]Wuhan National Laboratory for Optoelectronics, Huazhong University of Science and Technology, Wuhan 430074, China

[2]School of Optical and Electronic Information, Huazhong University of Science and Technology, Wuhan 430074, China

[3]Optics Valley Laboratory, Wuhan 430074, China

✉ e-mail: yuan_yu@hust.edu.cn; xlzhang@mail.hust.edu.cn

## Abstract

Multimode photonic integrated circuits enable ultralow-loss on-chip optical interconnects and microwave-photonic processing, yet waveguide bends dominate both chip footprint and excess loss. A high-performance multimode waveguide bend (MWB) must transmit the working mode with low loss while suppressing intermodal coupling, a requirement that forces existing schemes into a trade-off among bending radius, operating bandwidth, and fabrication tolerance. Here we establish, to the best of our knowledge, the first variational design framework for constant-width MWBs. By formulating bend optimization as a curvature-dependent figure-of-merit problem, we derive a universal necessary condition for adiabatic optimality. Across the entire bend, including its junctions with the input and output straight waveguides, the curvature profile must approach infinite differentiability. This condition explains the limitations of circular and Euler bends and leads to a smooth polynomial curvature (SPC) design strategy. Building on this framework, we propose an optimized SPC hybrid (SPCh) bend that combines a compact footprint, broadband operation, and high fabrication tolerance. On the 220 nm silicon-on-insulator platform, SPCh bends enable low-loss adiabatic transmission with a mode extinction ratio below −37 dB across 1500–1600 nm at an effective radius of 16 μm. This architecture provides over 22 dB higher mode suppression than a representative Euler bend at the same effective radius, while tolerating ±60 nm waveguide width fabrication deviations. The SPCh-based microring resonator reaches an intrinsic quality factor up to $7.53 \times 10^6$ with a free spectral range up to 100 GHz in an ultra-compact footprint on a standard silicon foundry process. This design principle is platform-independent, enabling compact multimode components for high-density optical interconnects and microwave-photonic systems.

---

## Introduction

Photonic integrated circuits underpin the next generation of dense optical interconnects [1–4], photonic-computing accelerators [5, 6], microwave-photonic processing [7–9], and on-chip quantum information [10–12]. As channel counts and subsystem complexity rise, waveguide bends increasingly govern signal routing across the chip layout and contribute substantially to both footprint usage and parasitic excess loss. Bend-related loss generally includes radiation loss, sidewall-scattering loss, and curvature-induced modal mismatch or intermodal coupling. Sidewall scattering is determined by etching roughness and scales with the optical field intensity at the core–cladding interface [13]. Curvature discontinuities or rapid curvature variations between adjacent sections can induce modal mismatch and intermodal coupling.

Widening the waveguide core reduces sidewall scattering by shifting the modal field away from the etched sidewalls. Multimode waveguides therefore achieve substantially lower propagation loss than their single-mode counterparts [14, 15]. As the core width increases, both the modal overlap with rough sidewalls and the effective-index sensitivity to width perturbations, $\partial n_{\mathrm{eff}}/\partial W$, decrease monotonically (see Section S1, Supporting Information for the full width dependence). Meanwhile, inverse-designed multimode passive components have matured on the silicon-on-insulator (SOI) platform [16, 17], increasing the demand for equally compact, low-loss interconnecting bends. The wide core, however, also supports the higher-order transverse-electric (TE) modes $TE_1$, $TE_2$, and beyond alongside the data-carrying fundamental $TE_0$. A multimode waveguide bend (MWB) must therefore route $TE_0$ with low loss in a compact footprint, while suppressing higher-order mode (HOM) excitation.

Existing MWB designs fall into two families. The first employs width modulation along the bend to suppress HOM excitation. Tapered transitions narrow the multimode core into a single-mode segment before bending and expand it back afterward [18, 19]. However, the narrow single-mode waveguide section partly offsets the low-scattering advantage of using a multimode core. Tao et al. [20] introduced an impressive “matched bend” combining an Euler segment, a constant-radius arc, and adiabatic width tapering. By exploiting time-reversal symmetry about the bend midpoint, it allows $TE_1$ excitation inside the bend yet cancels it at the output, reaching a high intrinsic quality factor of $9.6 \times 10^6$ at a compact effective radius of 18.2 μm. Because this cancellation relies on precise inter-modal phase matching, its operating bandwidth and fabrication tolerance leave room for further improvement. Width-modulated arcs, illustrated in Fig. 1a, follow similar principles and share comparable sensitivity limitations. Constant-width geometries, by contrast, avoid the phase-matching sensitivity of the width-modulated designs and remain a widely adopted baseline for MWB design. Accordingly, the second family, illustrated in Fig. 1b, maintains constant core width along the bend and optimizes only the curvature profile. The Euler bend [21–24] varies the curvature $\kappa$ linearly along the arc, providing continuous $\kappa$ at the junctions. Modified Euler bends with constant multimode width have reduced propagation losses to 0.10–0.15 dB/cm at effective radii of 29–49 μm on the 220 nm SOI platform [14]. Multi-segment Euler bends [25], together with Bezier and free-form profiles derived via brute-force or topology optimization [26–30], have further boosted performance. These approaches, however, rely on numerical search rather than analytical derivation. A unifying framework that derives the optimal curvature directly from fundamental mode-coupling principles has yet to be established.

Here, we formulate constant-width MWB design as a variational optimization problem. We define a figure of merit (FOM) that quantifies the loss and adiabatic transmission characteristics of the MWB, and then apply the Euler–Lagrange equation together with the Legendre necessary condition. The analysis yields a universal necessary condition for optimality: the optimal curvature profile of a constant-width MWB must approach infinite differentiability throughout the bend, including its junctions with the input and output straight waveguides. Circular arcs fail this condition because $\kappa$ is discontinuous at these junctions. Euler bends retain discontinuities in $\kappa'$ at the same junctions and at the bend midpoint. Guided by this principle, we introduce the smooth polynomial curvature (SPC) family and identify an SPC hybrid (SPCh) configuration that balances junction smoothness with the interior curvature gradient. The resulting constant-width SPC design concept is illustrated in Fig. 1c. Three-dimensional finite-difference time-domain (FDTD) simulations and microring resonator (MRR) experiments on the 220 nm SOI platform demonstrate three key results: (i) The SPCh bend enables low-loss adiabatic transmission with a mode extinction ratio (MER) below −37 dB across 1500–1600 nm at an effective radius of 16 μm, indicating over 22 dB higher mode suppression than a representative Euler bend at the same radius. The MER remains below −28 dB even when the operating window is broadened to 1300–1800 nm. (ii) The SPCh performance is preserved under ±60 nm waveguide-width deviation, indicating a large fabrication tolerance. (iii) The fabricated MRR reaches an intrinsic quality factor up to $7.53 \times 10^6$ and a free spectral range (FSR) up to 100 GHz in an ultra-compact footprint on a standard silicon multi-project wafer (MPW) process. These results collectively break the conventional trade-off between compact bend radius, broadband operation, and fabrication tolerance. Beyond SOI, this design principle transfers naturally to other multimode photonic platforms, providing a portable framework for compact, high-performance integrated photonics.

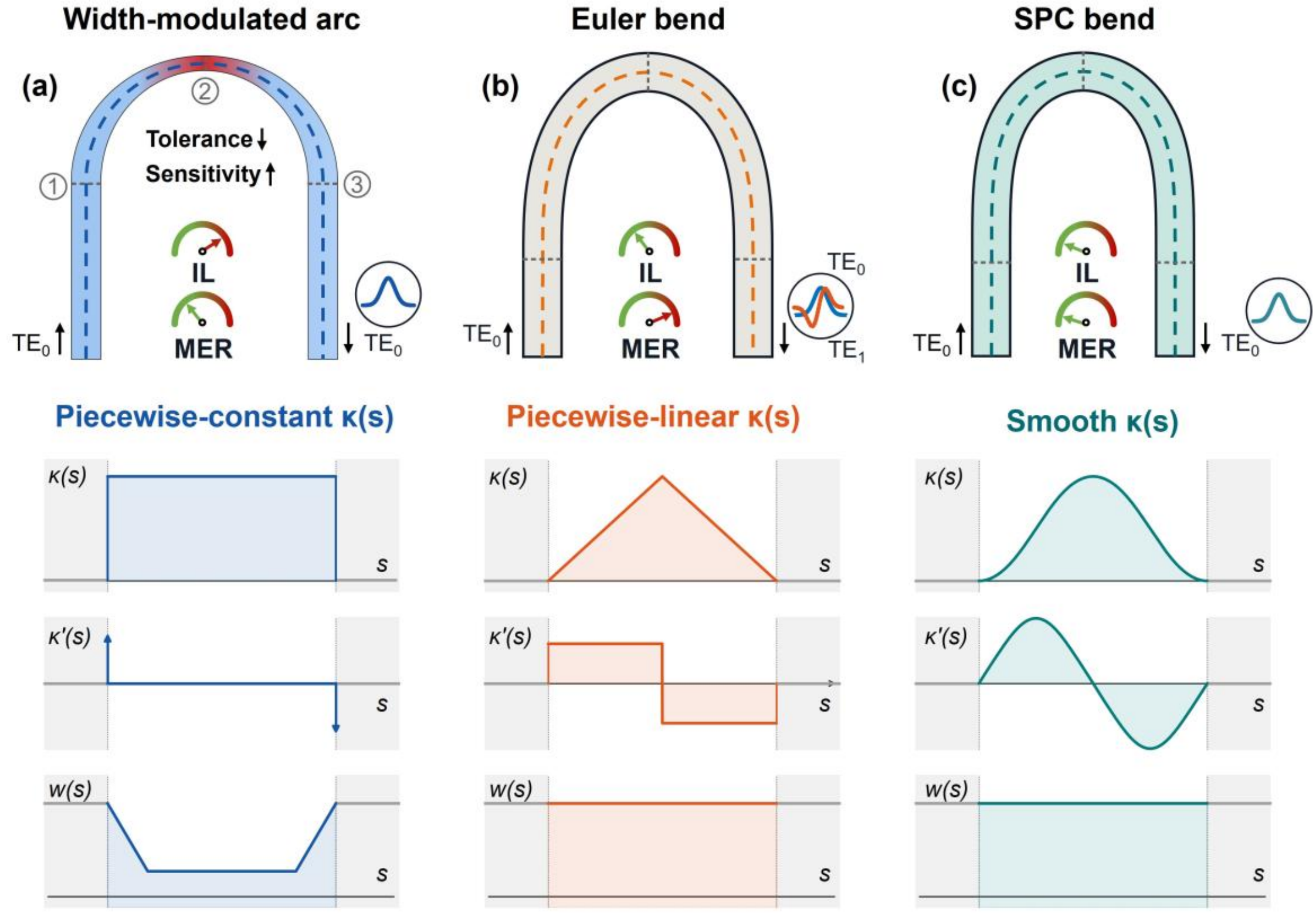

**Figure 1.** Concept of constant-width SPC MWBs. **a–c**, Three design strategies for MWBs: width-modulated arc (a), Euler bend (b), and SPC bend (c). The width-modulated arc varies the core width $w(s)$ along the bend, whereas the Euler and SPC bends hold the width constant and shape only the curvature profile $\kappa(s)$, both functions of the arc length $s$. The proposed SPC bend maintains curvature continuity, suppressing HOM excitation and enabling effective $TE_0$ transmission. The gauge dials in each panel qualitatively indicate the insertion loss (IL) and MER.

---

# Results

## Variational formulation and the infinite-differentiability condition

We consider two curvature-induced effects that limit low-loss adiabatic transmission in constant-width MWBs: local mode coupling and sidewall field enhancement. Fig. 2a illustrates the mode coupling caused by a curvature difference $\Delta\kappa = \kappa_{i+1} - \kappa_i$ between adjacent waveguide sections. The incident $TE_0$ is transmitted into the local $TE_0$ mode and can also couple to HOMs, with $TE_1$ shown as a representative HOM. The coupling efficiency into each mode is determined by the squared magnitude of the normalized overlap integral between the corresponding eigenmode fields on either side of the junction. The same local coupling calculation yields two related metrics: the HOM excitation ratio $\eta_{\text{HOM}}$, which measures the power coupled to HOMs, and the $TE_0$ mismatch loss $L_{\text{mis}}$, which measures the reduction in $TE_0$ transmission caused by mode mismatch. Fig. 2b shows the curvature-induced field shift at fixed waveguide width $W$. As $\kappa$ increases, the $TE_0$ field shifts toward the outer sidewall, increasing the normalized sidewall field intensity $\rho_{\mathrm{sw}}$.

To quantify these effects, we simulated the three corresponding metrics for a 1.6-µm-wide, 220-nm-thick SOI strip waveguide using Lumerical MODE. Fig. 2c–e show the simulated data and fitted curves. The HOM excitation ratio and $TE_0$ mismatch loss are fitted as quartic functions of $\Delta\kappa$ in Fig. 2c and d, whereas the normalized sidewall field intensity is fitted as a cubic function of the absolute curvature $\kappa$ in Fig. 2e.

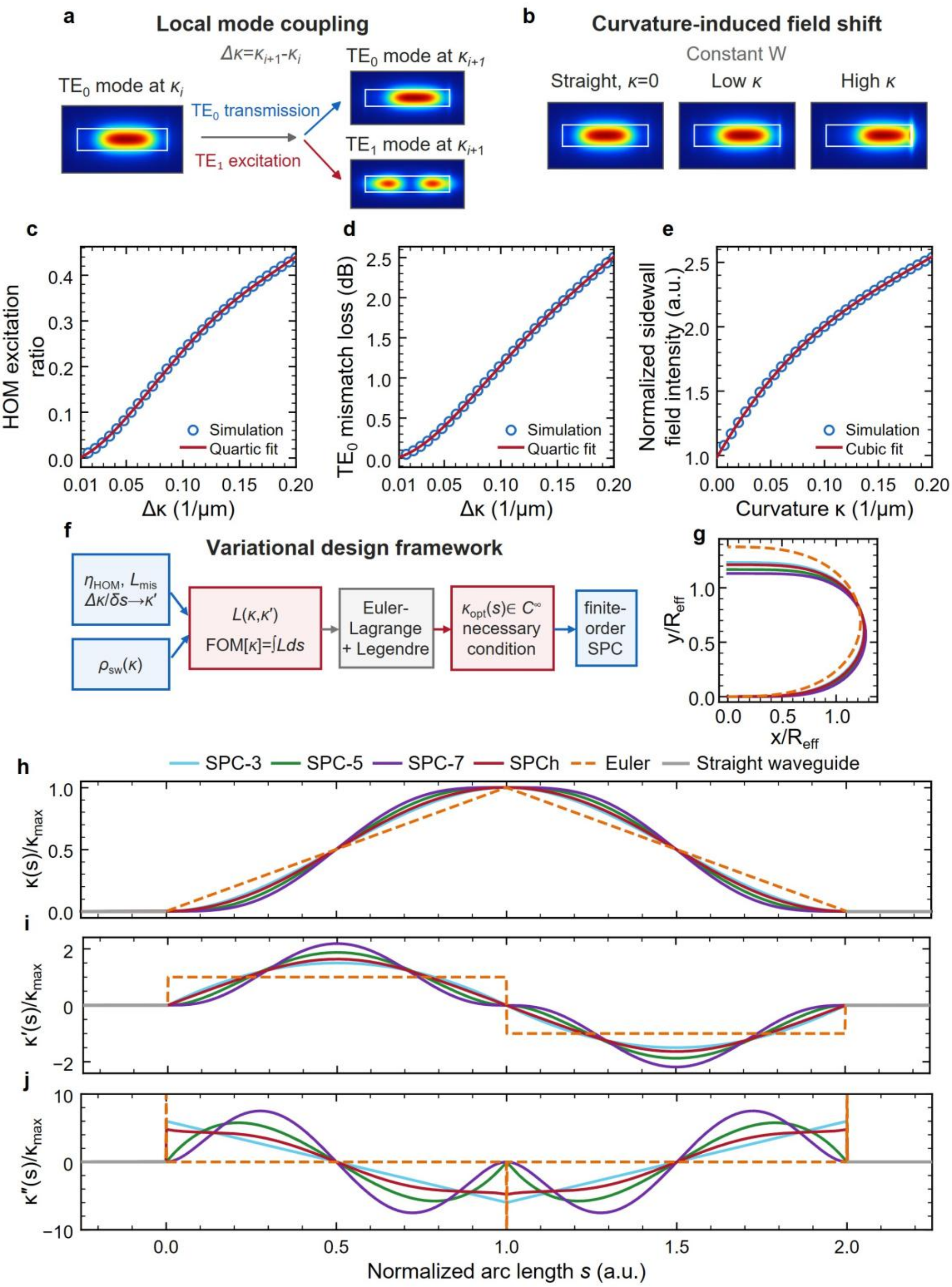

**Figure 2.** Design principle and variational framework for constant-width SPC MWBs. **a**, Schematic of local mode coupling between adjacent waveguide sections with curvatures $\kappa_i$ and $\kappa_{i+1}$, showing $TE_0$ transmission and excitation of the representative HOM $TE_1$ for $\Delta\kappa = \kappa_{i+1} - \kappa_i$. **b**, Schematic $TE_0$ field distributions in a straight section and constant-width bend sections at low and high $\kappa$. The outward field shift at high $\kappa$ is exaggerated for visual clarity. **c**, Simulated HOM excitation ratio $\eta_{\mathrm{HOM}}$ versus $\Delta\kappa$ and quartic fit. **d**, Simulated $TE_0$ mismatch loss $L_{\mathrm{mis}}$ versus $\Delta\kappa$ and quartic fit. **e**, Simulated normalized sidewall field intensity $\rho_{\mathrm{sw}}$ versus

absolute curvature $\kappa$ and cubic fit, referenced to the straight-waveguide sidewall field. The simulations in **c–e** use a 1.6-μm-wide, 220-nm-thick SOI strip waveguide with $SiO_2$ cladding and $TE_0$ excitation. **f**, Variational design framework that maps the $\Delta\kappa$-dependent $\eta_{\mathrm{HOM}}$ and $L_{\mathrm{mis}}$ to $\kappa'$ through $\Delta\kappa/\delta s \to \kappa'$, retains $\rho_{\mathrm{sw}}$ as a function of $\kappa$, and constructs $\mathscr{L}(\kappa, \kappa')$ and the FOM. The Euler–Lagrange equation and Legendre necessary condition yield $\kappa_{\mathrm{opt}}(s) \in C^{\infty}$ as a necessary condition, followed by a finite-order SPC approximation. $C^{\infty}$ denotes an infinitely differentiable optimal curvature profile. **g**, Centerline geometries reconstructed from the five curvature profiles at common $R_{\mathrm{eff}} = 2R_{\mathrm{min}}$ and common arc length. Their port openings differ because the curves are not individually rescaled. **h–j**, Normalized curvature $\kappa(s)/\kappa_{\mathrm{max}}$ (**h**), its first derivative $\kappa'(s)/\kappa_{\mathrm{max}}$ (**i**), and second derivative $\kappa''(s)/\kappa_{\mathrm{max}}$ (**j**) for five constant-width 180° bend designs, plotted as the SPC-3 (light blue), SPC-5 (green), SPC-7 (purple), SPCh (red), and Euler bend (orange dashed) curves along the normalized arc length $s$, where $s = 0.0$, 1.0, and 2.0 mark the bend entrance, bend midpoint, and bend exit, respectively, on the normalized arc-length coordinate. The gray flat segments at both ends mark the connecting straight waveguides.

The fitted mode-coupling metrics in Fig. 2c and d are functions of the curvature difference $\Delta\kappa$ between adjacent waveguide sections. As the arc-length interval $\delta s$ between adjacent sections approaches zero, $\Delta\kappa/\delta s$ approaches the curvature derivative $\kappa' = d\kappa/ds$. We therefore express these two metrics in terms of $\kappa'$, while $\rho_{\mathrm{sw}}$ remains a function of the local curvature $\kappa$. We combine the three fitted metrics in a Lagrangian density $\mathscr{L}(\kappa, \kappa')$ and define the FOM functional as its integral along the bend, $\mathrm{FOM}[\kappa] = \int \mathscr{L} ds$. The full expression is given by Equation (S5) in the Supporting Information. Within this variational formulation, the design of an ultralow-loss adiabatic MWB reduces to finding the curvature function $\kappa(s)$ that minimizes the FOM. Minimizing the FOM favors curvature profiles with low HOM excitation, low $TE_0$ mismatch loss, and low normalized sidewall field intensity. Applying the Euler–Lagrange equation together with the Legendre necessary condition yields a universal necessary condition for optimality: the optimal curvature must be infinitely differentiable throughout the bend, including its junctions with the input and output straight waveguides. Fig. 2f summarizes this variational design framework. The derivation in Sections S2 and S3 of the Supporting Information depends on the form of $\mathscr{L}$ rather than platform-specific coefficients. This necessary condition is therefore independent of platform, wavelength, and waveguide width.

Since a finite-order polynomial cannot satisfy the infinite-differentiability condition required at the junctions with the straight waveguides, we relax the infinite-differentiability requirement to high-order differentiability. We take an odd-order polynomial as the basic form of the curvature $\kappa(s)$ over the symmetric half-bend, with $s \in [0,1]$ denoting the normalized arc length (Section S4, Supporting Information):

$$\kappa_n(s) = \kappa_{\mathrm{max}} \sum_{k=0}^{n} a_k s^k. \quad (1)$$

Here, $\kappa_{\max} = 1/R_{\min}$ is the maximum curvature at the bend midpoint, $a_k$ are the polynomial coefficients fixed by the junction values $\kappa_n(0) = 0$ and $\kappa_n(1) = \kappa_{\max}$ together with the derivative conditions $\kappa_n^{(j)}(0) = \kappa_n^{(j)}(1) = 0$ for $j = 1, \dots, m$, and $n = 2m + 1$ with $m \in \mathbb{N}^+$. This odd order ensures the polynomial coefficients exactly match the paired boundary conditions for $m$-th-order curvature continuity, motivated by the symmetry of the boundary constraints and the full-rank requirement of the resulting algebraic system. The junction constraints uniquely fix the polynomial, whose smoothness increases with $m$ and approaches infinite differentiability in the limit (see Section S4, Supporting Information). We derive a closed-form general term for the entire SPC family,

$$\kappa_n(s) = \kappa_{\max} \frac{\int_0^s t^m (1-t)^m \, dt}{\int_0^1 t^m (1-t)^m \, dt} = \kappa_{\max} \, I_s(m+1, m+1), \tag{2}$$

where $I_s$ is the regularized incomplete Beta function (full derivation in Section S4, Supporting Information). Fig. 2g compares the corresponding bend centerlines at a common $R_{\text{eff}} = 2R_{\min}$, with the general definition of $R_{\text{eff}}$ given in Methods. Fig. 2h–j show their normalized curvature, first derivative, and second derivative along the complete bend. The explicit curvature functions of the five bends are provided in Section S4 of the Supporting Information. Fig. 2h shows that each profile changes from $\kappa = 0$ at the straight-waveguide junction to $\kappa_{\max}$ at the bend midpoint over each half-bend. Higher SPC orders match more curvature derivatives at these endpoints, making the curvature transition smoother there. Consequently, the curvature changes more rapidly in the interior of each half-bend, producing the progressively larger maximum $|\kappa'|$ shown in Fig. 2i. The improved boundary smoothness is therefore offset by a larger interior curvature gradient, so FOM$[\kappa]$ does not necessarily decrease with polynomial order.

We therefore introduce an SPC hybrid profile as a weighted combination of two normalized SPC basis profiles:

$$\kappa_{\text{hybrid}}(s) = w_3 \, \kappa_3(s) + w_7 \, \kappa_7(s), \tag{3}$$

where $w_3$ and $w_7$ are the weights of the SPC-3 and SPC-7 base profiles satisfying $w_3 + w_7 = 1$, and $\kappa_3(s)$ and $\kappa_7(s)$ are the basic SPC profiles obtained from Equation (1) with $n = 3$ and $n = 7$, respectively. The SPC-3 profile distributes the curvature change more broadly over each half-bend and therefore has a lower maximum $|\kappa'|$, but it provides less higher-order smoothness at the straight-waveguide junctions. The SPC-7 profile matches more curvature derivatives at these junctions, but shifts the curvature change into a narrower region within each half-bend and therefore has a larger maximum $|\kappa'|$. A weight sweep over $w_3 \in [0,1]$ (with $w_7 = 1 - w_3$) identifies $w_3 = 0.8, w_7 = 0.2$ as the joint optimum that balances the two effects (see Section S4, Supporting Information). We denote this 80:20 blend as SPCh, the specific optimized SPC hybrid implementation characterized in this work, giving $\kappa_{\mathrm{SPCh}}(s) = 0.8\,\kappa_3(s) + 0.2\,\kappa_7(s) = \kappa_{\max}(2.4s^2 - 1.6s^3 + 7s^4 - 16.8s^5 + 14s^6 - 4s^7)$. The resulting SPCh profile balances the gradual curvature variation of SPC-3 against the higher-order junction smoothness of SPC-7. As Fig. 2i and j show, this balance provides a smooth straight-to-bend transition while keeping the maximum $|\kappa'|$ below that of SPC-7.

## Compact-radius performance

To identify the optimal SPC order configuration, we compared five 180° bend curvature profiles (SPC-3, SPC-5, SPC-7, SPCh, and Euler bend) at the same compact effective bend radii. We selected effective bend radii $R_{\mathrm{eff}}$ = 16 μm at $W$ = 1.6 μm and $R_{\mathrm{eff}}$ = 30 μm at $W$ = 2.0 μm, approximately half the typical Euler bend radius used in prior modified-Euler microring designs on the 220 nm SOI platform [14, 31]. The key performance metrics are the maximum IL ($\mathrm{IL_{max}}$) and maximum MER ($\mathrm{MER_{max}}$), each taken as the worst-case value over the 1500–1600 nm operating band (definition in Methods).

Fig. 3a plots all ten configurations on the $\mathrm{IL_{max}}$–$\mathrm{MER_{max}}$ plane. At $W$ = 1.6 μm, SPCh achieves $\mathrm{IL_{max}}$ = 0.004 dB and $\mathrm{MER_{max}}$ = −37.5 dB, whereas the $\mathrm{IL_{max}}$ and $\mathrm{MER_{max}}$ of the Euler bend at the same $R_{\mathrm{eff}}$ are 0.144 dB and −14.8 dB, respectively. At $W$ = 2.0 μm, SPCh maintains an equally low $\mathrm{IL_{max}}$ and improves $\mathrm{MER_{max}}$ to −39.1 dB, whereas the $\mathrm{IL_{max}}$ and $\mathrm{MER_{max}}$ of the Euler bend are 0.046 dB and −20.0 dB, respectively. At both widths, SPCh combines low IL with strong mode suppression and outperforms the Euler bend on both metrics, confirming it as the optimal configuration among the SPC orders.

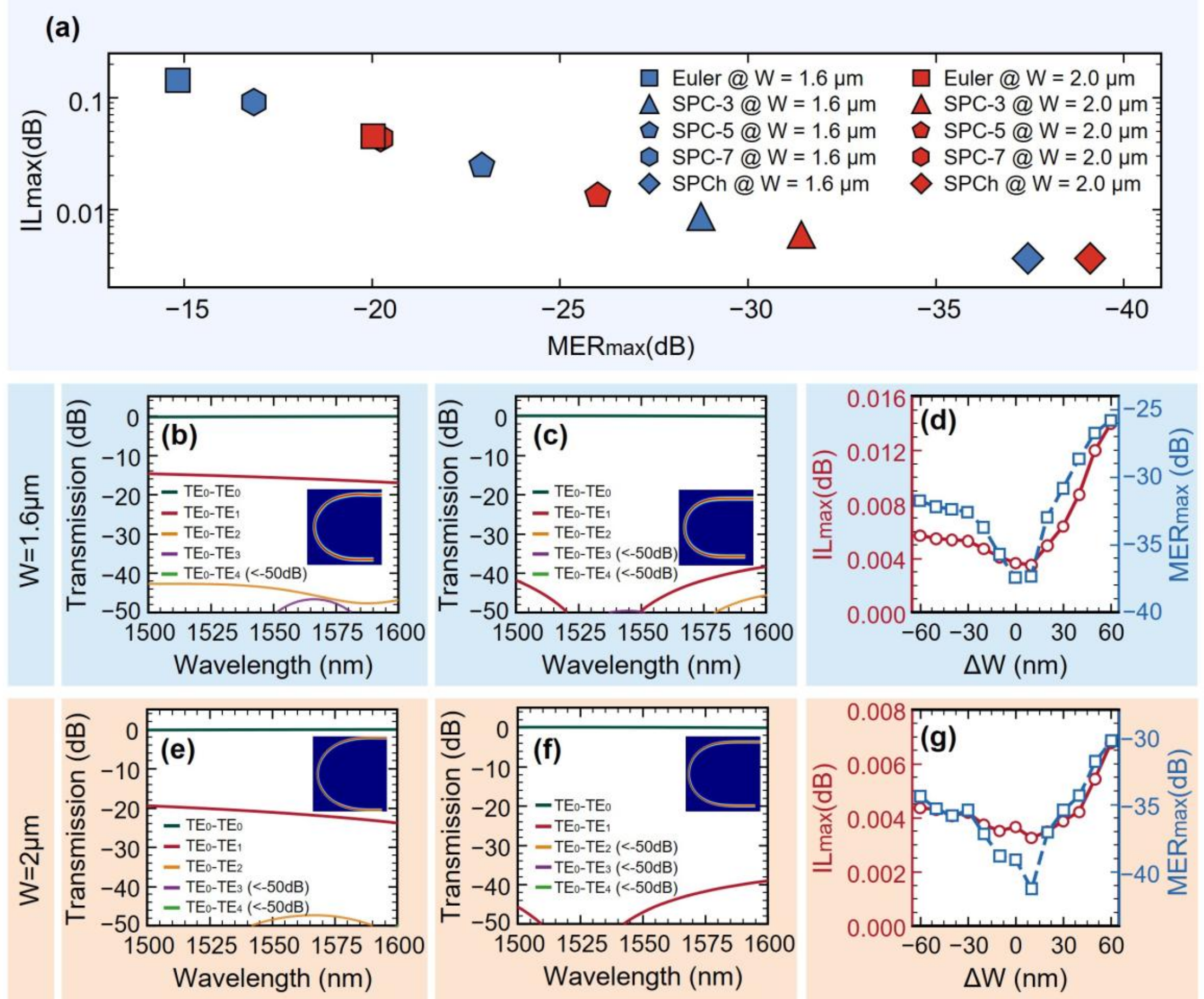

**Figure 3.** Compact-radius performance of constant-width SPC bends. **a**, Comparison of five curvature types (SPC-3, SPC-5, SPC-7, SPCh, and Euler bend) at two waveguide widths, showing $IL_{max}$ versus $MER_{max}$ across 1500–1600 nm at the compact operating points $R_{eff}$ = 16 μm for $W$ = 1.6 μm (blue) and $R_{eff}$ = 30 μm for $W$ = 2.0 μm (red). Marker shape encodes curvature type and fill color encodes waveguide width, so each (shape, color) pair maps one-to-one to a single (curvature, width) device. **b**, Transmission spectrum of the 180° Euler bend at $W$ = 1.6 μm, $R_{eff}$ = 16 μm, with $TE_0$ launched and the output decomposed onto $TE_0$–$TE_4$ across 1500–1600 nm. **c**, Transmission spectrum of the 180° SPCh bend at $W$ = 1.6 μm, $R_{eff}$ = 16 μm. **d**, Width tolerance of the SPCh bend at $W$ = 1.6 μm, $R_{eff}$ = 16 μm: $IL_{max}$ (red, left axis) and $MER_{max}$ (blue, right axis) versus width deviation $\Delta W$. **e**, Transmission spectrum of the 180° Euler bend at $W$ = 2.0 μm, $R_{eff}$ = 30 μm. **f**, Transmission spectrum of the 180° SPCh bend at $W$ = 2.0 μm, $R_{eff}$ = 30 μm. Insets in **b**, **c**, **e**, **f** show the simulated light propagation in the designed waveguide. **g**, Width tolerance of the SPCh bend at $W$ = 2.0 μm, $R_{eff}$ = 30 μm: $IL_{max}$ (red, left axis) and $MER_{max}$ (blue, right axis) versus width deviation $\Delta W$.

Two conclusions emerge from Fig. 3. First, even the SPC-3 bend outperforms the Euler bend at both widths, confirming that any smooth polynomial profile satisfying the boundary conditions of Equation (1) surpasses the $\kappa'$-discontinuous Euler bend. This advantage arises because SPC-3 enforces continuity of both $\kappa$ and $\kappa'$ at the junctions between the bend and the straight waveguide, whereas the Euler bend retains a finite $\kappa'$ discontinuity at the same junctions. Second, the bend performance, quantified by $\mathrm{IL}_{\max}$ and $\mathrm{MER}_{\max}$, does not improve monotonically with polynomial order: the SPC-5 and SPC-7 bends both underperform the SPC-3 bend, and the SPC-7 bend approaches the Euler bend baseline. As Fig. 2i shows, increasing the SPC order shifts the curvature change away from the ends of each half-bend and into a narrower interior region, raising the maximum $|\kappa'|$. This offsets the benefit of matching more curvature derivatives at the straight-waveguide junctions. The 80:20 SPCh profile balances these effects.

The transmission spectra in Fig. 3b, c, e, and f confirm that the SPCh mode-suppression advantage over the Euler bend holds across the full 1500–1600 nm band. SPCh exhibits uniform $TE_0$ transmission at both waveguide widths, with excitation of all HOMs suppressed below −37 dB. By contrast, the Euler bend at identical $R_{\mathrm{eff}}$ shows $TE_1$ excitation levels as high as −15 dB.

Fig. 3d and g show that, at the compact radii ($R_{\mathrm{eff}}$ = 16 and 30 µm), the SPCh performance remains tolerant to lithographic error. Across the full ±60 nm range, the IL and MER stay below 0.014 dB and −25.8 dB at $W$ = 1.6 µm, respectively. At $W$ = 2.0 µm, the IL and MER stay below 0.007 dB and −30.3 dB, respectively. Even with a +60 nm width error, SPCh outperforms the Euler bend at zero width error for the $W$ = 1.6 µm design, with an $\mathrm{IL}_{\max}$ of 0.014 dB versus 0.144 dB and an $\mathrm{MER}_{\max}$ of −25.8 dB versus −14.8 dB. SPCh bends suppress HOM excitation at the bend entrance, so small fabrication-induced width perturbations do not degrade the broadband performance.

## Bend-radius scaling

To confirm that SPCh's advantages are intrinsic to the design rather than a coincidence at a single radius or wavelength, we swept the effective radius $R_{\mathrm{eff}}$ from 16 to 32 µm at $W$ = 1.6 µm and from 30 to 60 µm at $W$ = 2.0 µm. Figs. 4a and d plot $\mathrm{IL}_{\max}$ versus effective radius for $W$ = 1.6 µm and $W$ = 2.0 µm, respectively. Figs. 4b and e plot $\mathrm{MER}_{\max}$ for the SPCh, Euler, and circular-arc bends at $W$ = 1.6 µm and $W$ = 2.0 µm, respectively.

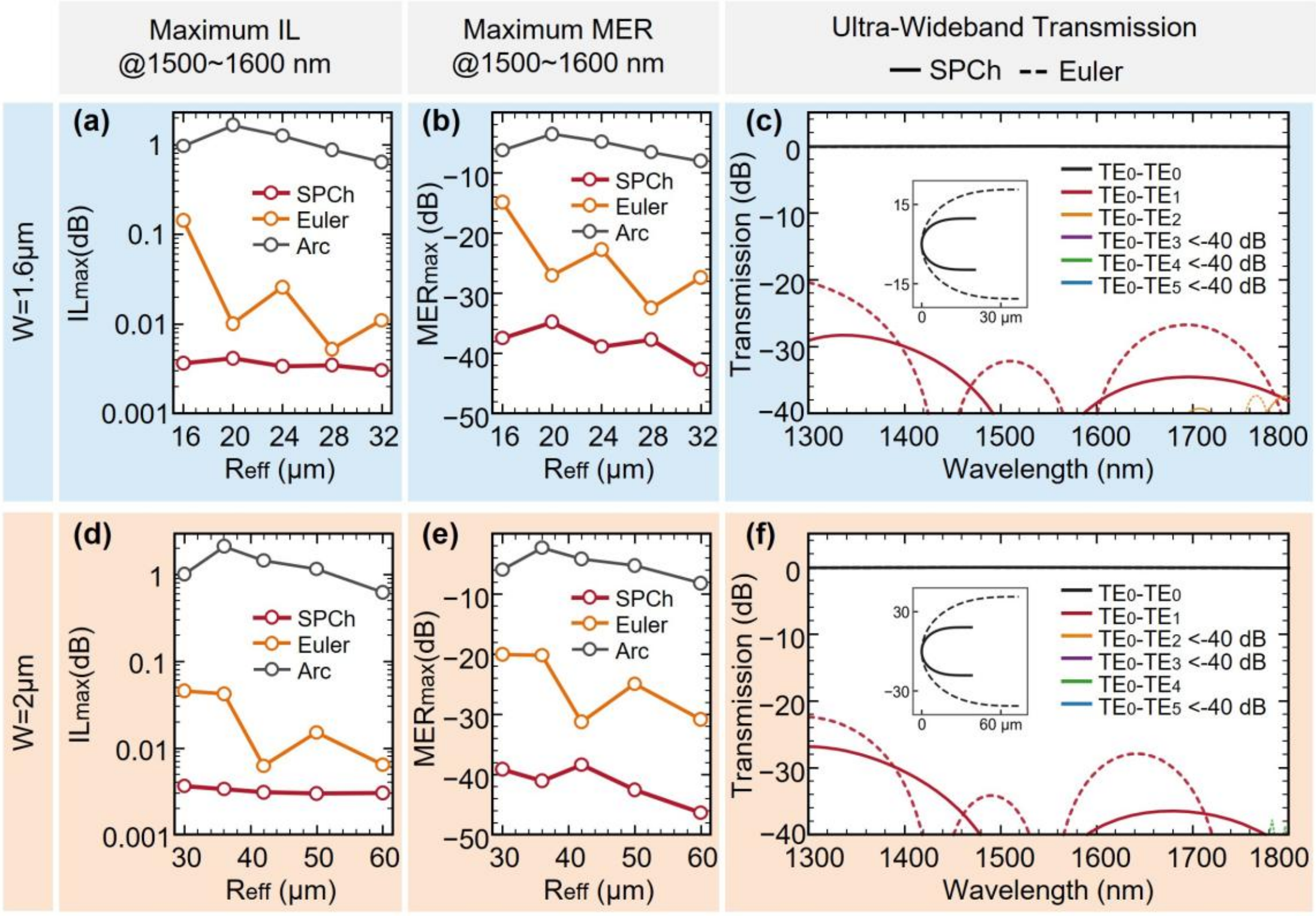

**Figure 4.** Bend-radius scaling and ultra-wideband transmission of SPCh, Euler bend and circular-arc 180° bends. **a**, $\mathrm{IL}_{\max}$ versus effective bend radius $R_{\mathrm{eff}}$ at $W$ = 1.6 μm across 1500–1600 nm, for SPCh (red), Euler bend (orange), and circular arc (gray). **b**, $\mathrm{MER}_{\max}$ versus $R_{\mathrm{eff}}$ at $W$ = 1.6 μm. **c**, Transmission spectra across 1300–1800 nm of the compact SPCh bend ($R_{\mathrm{eff}}$ = 16 μm, solid) and a relaxed Euler bend ($R_{\mathrm{eff}}$ = 30 μm, dashed) at $W$ = 1.6 μm. **d**, $\mathrm{IL}_{\max}$ versus $R_{\mathrm{eff}}$ at $W$ = 2.0 μm. **e**, $\mathrm{MER}_{\max}$ versus $R_{\mathrm{eff}}$ at $W$ = 2.0 μm. **f**, Transmission spectra of the compact SPCh bend ($R_{\mathrm{eff}}$ = 30 μm, solid) and a relaxed Euler bend ($R_{\mathrm{eff}}$ = 60 μm, dashed) at $W$ = 2.0 μm. The insets in **c** and **f** show the two bend outlines to scale.

SPCh holds $\mathrm{IL}_{\max}$ within 0.003–0.004 dB across the full radius scan at both widths. $\mathrm{MER}_{\max}$ remains below −34.8 dB at all bend radii, deepening to −46.3 dB at the largest bend radius of $R_{\mathrm{eff}}$ = 60 μm. By contrast, the Euler bend oscillates non-monotonically with radius, spanning 0.005 to 0.144 dB in $\mathrm{IL}_{\max}$ and −14.8 to −32.4 dB in $\mathrm{MER}_{\max}$. At $R_{\mathrm{eff}}$ = 28 μm, its $\mathrm{IL}_{\max}$ and $\mathrm{MER}_{\max}$ coincidentally reach 0.005 dB and −32.4 dB, but degrade to 0.011 dB and −27.4 dB at 32 μm. These oscillations arise from mode beating between $TE_0$ and the excited $TE_1$, governed by the bend-length-to-beat-length ratio rather than by design optimality. The circular-arc baseline shows significantly higher $\mathrm{IL}_{\max}$ and weaker mode suppression than SPCh and Euler bends across the entire scan.

Figs. 4c and f compare the transmission spectra of compact SPCh bends with those of Euler bends at twice the effective radius across the 1300–1800 nm wavelength band (covering O, E, S, C, L, and U telecom bands). Across this band, all four traces show similar $TE_0$ throughput, with $\mathrm{IL}_{\max} \leq$ 0.092 dB. Adiabatic transmission, however, clearly differentiates the two designs. At $W$ = 1.6 μm, the compact SPCh holds $MER_{max}$ at −28.2 dB and surpasses the relaxed Euler bend by 7.9 dB. At $W$ = 2.0 µm, it reaches −26.7 dB and outperforms the relaxed Euler bend by 4.4 dB. The compact SPCh thus preserves the ultra-broadband insertion-loss budget of the Euler bend at twice the effective radius while maintaining substantially stronger mode suppression across the same span.

The SPCh advantage extends to 90° bends at both compact and relaxed radii. At $W$ = 1.6 µm and $R_{eff}$ = 20 µm, SPCh maintains $MER_{max}$ at −27.1 dB across 1500–1600 nm, 15.1 dB lower than that of the Euler bend at identical radius. At $W$ = 2.0 µm and $R_{eff}$ = 24 µm, the gap widens to 21.7 dB, with SPCh at −31.0 dB versus −9.4 dB for the Euler bend. SPCh also maintains $IL_{max}$ below 0.012 dB at both compact 90° points, whereas Euler bends reach 0.274 dB at $W$ = 1.6 µm and 0.487 dB at $W$ = 2.0 µm (see Section S5, Supporting Information).

## Experimental verification

We fabricated SPCh-bend MRRs on a standard 220 nm SOI MPW process at Advanced Micro Foundry (AMF, Singapore). Fig. 5a–c show microscope images of the three fabricated ultrahigh-$Q$ racetrack resonators, with compact footprints of 0.39 × 0.02 mm² (Device A), 0.78 × 0.04 mm² (Device B), and 0.43 × 0.04 mm² (Device C), respectively. Each racetrack resonator consists of two 180° SPCh bends connected by two straight waveguide segments and is side-coupled to a bus waveguide through a symmetric straight-waveguide directional coupler. These bends operate at $R_{eff}$ = 16 µm for $W$ = 1.6 µm and $R_{eff}$ = 30 µm for $W$ = 2.0 µm.

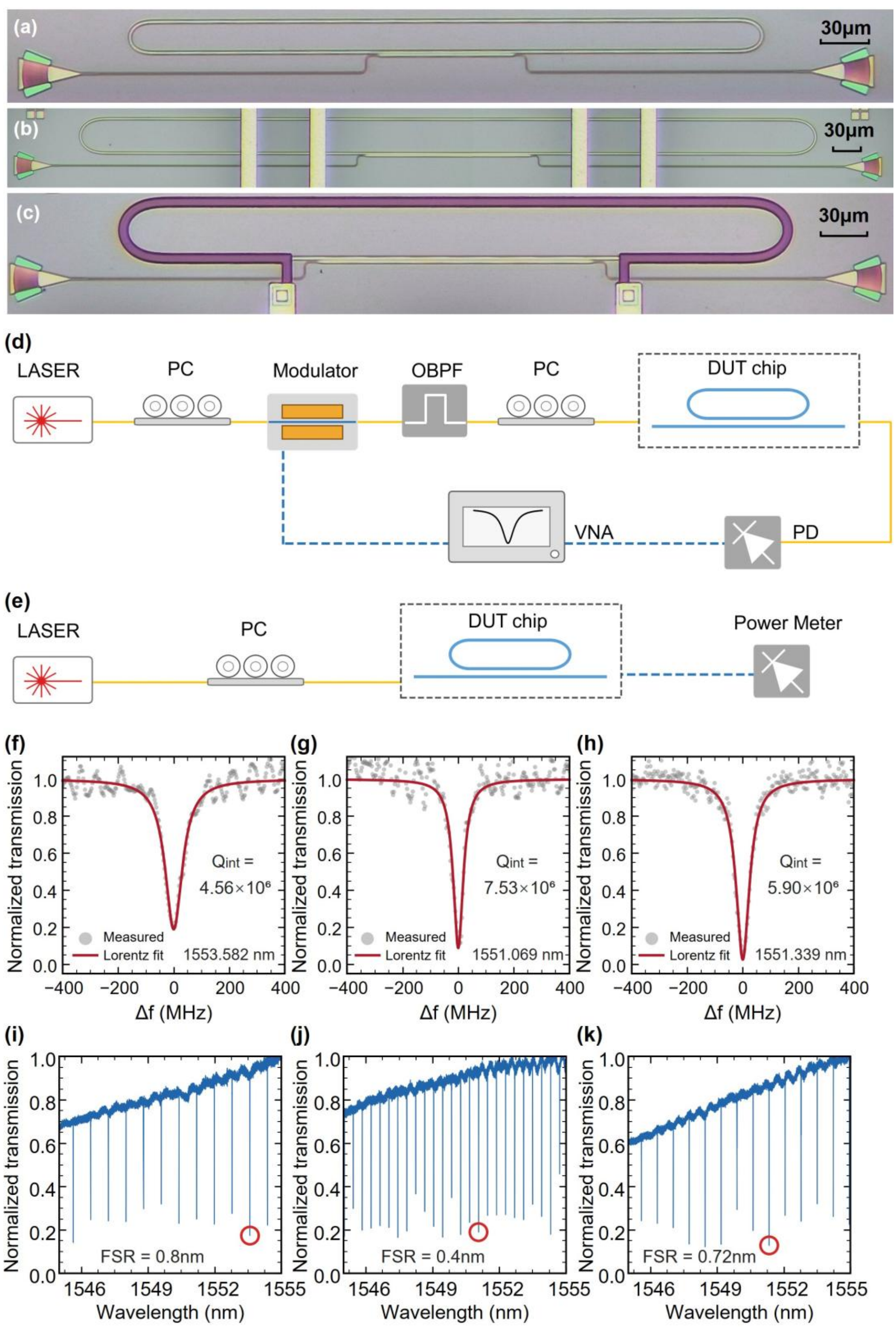

**Figure 5.** Experimental $Q$ validation of compact SPCh-based MRRs on 220 nm SOI. **a–c**, Optical micrographs of the three fabricated SPCh-MRR racetracks: Device A (a, $W$ = 1.6 μm, $R_{\text{eff}}$ = 16 μm, perimeter $L$ = 800.5 μm), Device B (b, $W$ = 2.0 μm, $R_{\text{eff}}$ = 30 μm, $L$ = 1588.5 μm), and Device C (c, $W$ = 2.0 μm, $R_{\text{eff}}$ = 30 μm, $L$ = 888.5 μm). Each ring is closed by two 180° SPCh bends and two straight waveguide segments. Microheaters above the resonator are visible in (c) and provide thermo-optic tuning. **d**, Phase-modulated microwave-photonic

link with vector network analyzer (VNA) sweep for high-resolution $Q$ value measurement. **e**, Broadband swept-laser transmission setup. **f–h**, Lorentzian fits of single resonances of the three devices, yielding intrinsic quality factors $Q_{int}$ = 4.56 × $10^6$ at 1553.582 nm (f, Device A), 7.53 × $10^6$ at 1551.069 nm (g, Device B), and 5.90 × $10^6$ at 1551.339 nm (h, Device C). **i–k**, Broadband transmission across 1545–1555 nm for Device A (i), Device B (j), and Device C (k). The red ring marks the resonance used for the Lorentzian fit in (f–h).

Two complementary setups characterized device performance. The phase-modulated microwave-photonic link (Fig. 5d) probed a single resonance with sub-MHz frequency resolution at low optical power, avoiding the thermal nonlinearity that distorts high-power scans of high-$Q$ cavities. The swept-laser system (Fig. 5e) traded resolution for a broad wavelength scan, mapping the resonance spectrum across tens of nanometres. Both setups used a manual fiber polarization controller to align the input light with the TE polarization (see Methods).

Figs. 5f–h show Lorentzian fits of single resonances within ±400 MHz of their centers, recorded with the phase-modulated microwave-photonic link (Fig. 5d). The fits yield intrinsic quality factors $Q_{int}$ = 4.56 × $10^6$ at 1553.582 nm, 7.53 × $10^6$ at 1551.069 nm, and 5.90 × $10^6$ at 1551.339 nm for Devices A, B, and C, respectively. Device B reaches the lowest propagation loss of 0.088 dB/cm among the fabricated devices (see Methods).

Figs. 5i–k show the broadband transmission of the same devices across 1545–1555 nm, recorded with the swept-laser transmission setup (Fig. 5e). The periodic resonance pattern is uniform across more than 10 nm, with no mode-dependent dips or anti-crossings. Only $TE_0$ resonances appear, directly confirming that SPCh bends support stable single-mode transmission across the measurement range. Combined with the simulated 1500–1600 nm flatness (Figs. 3c and f), this validates broadband single-mode operation.

As listed in Table 1, the three devices cover the two widths and three perimeters: $L$ = 800.5 μm (Device A), 1588.5 μm (Device B), and 888.5 μm (Device C). The FSR is calculated as $\lambda^2/(n_g L)$, where $\lambda$ is the operating wavelength, $n_g$ the $TE_0$ group index obtained from a 2D finite-difference eigenmode (FDE) solver, and $L$ the resonator round-trip length. Calculated FSRs (0.80, 0.40, and 0.72 nm at $\lambda$ = 1550 nm) match the measured values to within the 0.1 pm step of the swept-laser system, as listed in Table 1.

**Table 1. Summary of the three fabricated SPCh-MRR devices on 220 nm SOI.**

| Device | $W$ (μm) | $R_{eff}$ (μm) | FSR (nm) | FWHM[a] (MHz) | $Q_{int}$ (×$10^6$) | α (dB/cm) |
|---|---|---|---|---|---|---|
| A | 1.6 | 16 | 0.80 | 59.0 | 4.56 | 0.145 |
| B | 2.0 | 30 | 0.40 | 39.6 | **7.53** | **0.088** |

| Device | $W$ (μm) | $R_{\mathrm{eff}}$ (μm) | FSR (nm) | FWHM[a] (MHz) | $Q_{\mathrm{int}}$ ($\times 10^6$) | α (dB/cm) |
|---|---|---|---|---|---|---|
| C | 2.0 | 30 | 0.72 | 56.7 | 5.90 | 0.112 |

[a] FWHM, full width at half maximum of the resonance.

The ultrahigh $Q$ and large FSR enabled by SPCh bends together translate into narrow-bandwidth, wide-tuning-range microwave-photonic filtering capability. To demonstrate this, we integrated Device C ($W$ = 2.0 μm) into a tunable single-passband microwave-photonic filter (MPF) [32–34]. As visible in Fig. 5c, microheaters deposited over the individual MRR waveguides provide thermo-optic tuning of the resonance wavelength, which sets the passband center frequency of the MPF. The MPF achieves a 3-dB radio-frequency bandwidth of 76 MHz across a tuning range exceeding 40 GHz (see Section S6, Supporting Information).

---

## Discussion and Conclusion

### Performance comparison

Table 2 benchmarks our devices against prior silicon MRRs reporting intrinsic $Q$ on the 220 nm SOI platform, covering modified Euler bends [14, 31], matched bends [20], Bezier free-form curves [30], and other curvature designs across compact and relaxed effective radii.

**Table 2. Comparison of wide MWB designs for high-Q silicon MRRs on 220 nm SOI.**

| Ref. | $W$ (μm) | Bend type | $R_{eff}$ (μm) | $Q_{int}$ ($\times10^6$) | α (dB/cm) | FSR (nm) | BW[a] (nm) | Process |
|---|---|---|---|---|---|---|---|---|
| [31] | 1.6 | Modified Euler | 29 | 2.3 | 0.3 | 0.9 | 100[b] | MPW |
| [35] | 1.2/1.5 | Circular arc | 800 | ~2.4[c] | 0.28 | 0.127 | – | RP+Ox |
| [36] | 0.9 to 11 | Adiabatic width taper | 50 | 2.0 | 0.28 | 2.0 | – | LOCOS |
| [19] | 0.5 to 2 | Circular arc | 20 | 3.16 | 0.21 | 0.208 | – | MPW |
| [20] | 2 to 1.535 | Phase-matched Euler-arc | 18.2 | 9.6 | ~0.069[c] | 0.92 | 24[b] | MPW |
| [14] | 1.6 | Modified Euler | ~29 | ~4.4[c] | 0.15 | 0.325 | 100[b] | MPW |
| [14] | 2.0 | Modified Euler | ~49 | ~6.6[c] | 0.10 | 0.325 | 100[b′] | MPW |
| [14] | 3.0 | Modified Euler | 115 | ~10.2[c] | 0.065 | 0.325 | 100[b′] | MPW |
| [37] | 0.5/3 | Circular arc | ~5900[d] | 3.6 | ~0.18[c] | 0.017 | – | MPW |
| [30] | 0.45 to 1.6 | Bezier free-form | 20 | ~2.8[c] | 0.24 | 1.0 | 40[b] | MPW |
| **Ours (Device A)** | **1.6** | **SPCh** | **16** | **4.56** | **0.145** | **0.80** | **397[b]** | **MPW** |
| **Ours (Device B)** | **2.0** | **SPCh** | **30** | **7.53** | **0.088** | **0.40** | **407[b]** | **MPW** |
| **Ours (Device C)** | **2.0** | **SPCh** | **30** | **5.90** | **0.112** | **0.72** | **407[b]** | **MPW** |

[a] BW, optical operating bandwidth.

[b] Wavelength range with inter-mode crosstalk below −30 dB.

[b′] Wavelength range with inter-mode crosstalk below −25 dB.

[c] Marked $Q_{int}$ or $\alpha$ back-derived from the parameters reported in the article via $\alpha = 2\pi n_g/(Q_{int}\lambda) \cdot 4.343$ at $\lambda$ = 1550 nm, where $n_g$ is the waveguide group index.

[d] Full-ring equivalent radius, L/(2π).

Our SPCh devices concurrently integrate three key attributes: compact effective radius, broadband operation, and large fabrication tolerance. Device A operates at $R_{eff}$ = 16 μm ($W$ = 1.6 μm), and Device B at $R_{eff}$ = 30 μm ($W$ = 2.0 μm). Both achieve competitive intrinsic quality factors at these compact radii on the 220 nm SOI platform: 4.56 × $10^6$ at $R_{eff}$ = 16 μm and 7.53 × $10^6$ at $R_{eff}$ = 30 μm. Comparable intrinsic $Q$ on this platform has previously required either phase-matched bends, which narrow the operating band and fabrication tolerance [20], or substantially larger bend radii [14, 31]. Furthermore, the operating bandwidth reaches 397 nm and 407 nm at $W$ = 1.6 and 2.0 μm, respectively, with width tolerance reaching ±60 nm. The simultaneous combination of these three properties stems from the three design choices of Fig. 1: wide multimode core, constant width along the bend, and smooth high-order polynomial curvature profile.

## Platform portability

The infinite-differentiability necessary condition depends only on the polynomial structure of the Lagrangian density, and should transfer unmodified to other multimode platforms. To verify this, we repeated the mode-decomposition simulation on x-cut lithium niobate on insulator (LNOI) [38, 39] at $R_{\mathrm{eff}}$ = 100 μm for both $W$ = 1.5 μm and $W$ = 2.5 μm (see Section S7, Supporting Information). SPCh maintains $\mathrm{IL}_{\mathrm{max}}$ below 0.006 dB and $\mathrm{MER}_{\mathrm{max}}$ at −36.3 dB and −37.5 dB for $W$ = 1.5 and 2.5 μm, respectively. The Euler bend at identical radius reaches only −29.7 dB and −27.8 dB, corresponding to 6.6 dB and 9.7 dB lower mode suppression than SPCh. The SPCh-over-Euler advantage mirrors the SOI case despite lower core-index contrast and distinct HOM structure in LNOI. The same construction should extend to other low-loss multimode platforms such as silicon nitride [42–45].

## Outlook and conclusion

The proposed design principle serves as a drop-in replacement for any 90° or 180° bend in compact integrated photonic circuits, including spiral delay lines, multimode arrayed-waveguide gratings, and racetrack microring filters [40, 41]. Its effectiveness for 90° bends is confirmed in Section S5 of the Supporting Information. The S-bend extension follows from the symmetric construction detailed in Section S8 of the Supporting Information. Moreover, the same smoothness argument applies to directional couplers, as both arms inherit a smooth $\kappa(s)$. As a system-level application, we further demonstrated a tunable single-passband MPF (Section S6, Supporting Information).

In summary, we recast multimode-bend design as a variational problem and derive a single governing principle: the optimal curvature must approach infinite differentiability at every point along the bend. This principle explains why circular and Euler bends are suboptimal and yields the constant-width SPCh bend on a wide multimode core. On the 220 nm SOI platform, SPCh maintains a mode extinction ratio below −37 dB across 1500–1600 nm at a 16 μm effective radius, over 22 dB beyond a comparable Euler bend. Its single-mode adiabatic transmission is preserved under ±60 nm width deviations. The SPCh-based MRRs further reach an intrinsic quality factor up to $7.53 \times 10^6$ and a free spectral range up to 100 GHz on a standard silicon foundry process. The SPCh bend thus breaks the long-standing trade-off between compact bend radius, broadband operation, and fabrication tolerance in a single, foundry-ready geometry. Because the principle depends on the functional form of the loss model rather than material-specific parameters, it can be extended to other multimode photonic platforms, including silicon nitride and LNOI. Across these platforms, SPCh bends serve as compact, portable components for high-density optical interconnects and microwave-photonic systems.

# Methods

## Numerical simulation

The 3D FDTD simulations are performed via commercial software (Lumerical FDTD Solutions). Each model comprises an input multimode straight section, a 180° MWB, and an output multimode straight section, with the mode source and the field monitor placed on the input and output straight sections, respectively. Transmitted fields at the output port are decomposed onto the straight-waveguide eigenmodes, yielding mode-resolved transmission spectra $T_i(\lambda)$ plotted in Figs. 3 and 4. We define $\mathrm{IL_{max}} = -10\log_{10}[\min_\lambda T_{\mathrm{TE_0}}(\lambda)]$ and $\mathrm{MER_{max}} = 10\log_{10}[\max_\lambda(\max_{i\geq 1}(T_i(\lambda)/T_{\mathrm{TE_0}}(\lambda)))]$ over the 1500–1600 nm operating band, so that each metric reports the worst-wavelength performance across the band.

## Effective radius

For all 180° bends in this work, the effective radius $R_{\mathrm{eff}}$ is defined as the radius of a 180° constant-radius arc with the same arc length as the SPCh or Euler bend. Integrating the SPCh and Euler curvature profiles yields a 180° bend arc length of $2\pi/(1/R_{\min} + 1/R_{\max})$, which equals the arc length $\pi \cdot R_{\mathrm{eff}}$ of an equivalent 180° constant-radius bend. Therefore, $R_{\mathrm{eff}} = 2/(1/R_{\min} + 1/R_{\max})$. For SPCh and Euler bends in this work, the curvature ranges from $\kappa = 0$ ($R_{\max} = \infty$) at the junctions to $\kappa = 1/R_{\min}$ at the midpoint, so $R_{\mathrm{eff}} = 2\ R_{\min}$. For the constant-radius arc ($R_{\min} = R_{\max}$), $R_{\mathrm{eff}} = R_{\min}$.

## Microwave-photonic Q-factor measurement

Intrinsic $Q$ factors are measured using the phase-modulated microwave-photonic link shown in Fig. 5d. A continuous-wave carrier from a single-frequency tunable laser (NKT Photonics, Basik E15) passes through a fiber polarization controller into a lithium niobate phase modulator (Covega, Mach-40) driven by the swept radio-frequency tone of the VNA (Anritsu MS4647B). A tunable optical band-pass filter rejects one first-order sideband, and the remaining single-sideband signal is aligned to TE and coupled to the chip via tapered single-mode fibers (OMT, APC-TJ-1M) and on-chip grating couplers. A high-speed photodetector (SHF AG, Berlin) recovers the transmitted signal and feeds it to the VNA's S-parameter input, sampling the cavity transfer function around each resonance. No optical amplifier is used between the band-pass filter and chip, as SPCh-MRR resonance depths (7–16 dB) provide sufficient signal-to-noise ratio. Three single resonances at 1551.069, 1551.339, and 1553.582 nm are fitted to a Lorentzian within ±400 MHz of their centers to extract the loaded linewidth.

## Broadband transmission measurement

The broadband transmission spectra in Figs. 5i–k are recorded using the automated swept-laser system shown in Fig. 5e, which comprises a tunable continuous-wave laser (Santec, TSL-510) and a wavelength-synchronized optical power meter (Santec, MPM-210). The wavelength sweep covers 1545–1555 nm with a step of 0.1 pm and a sweep speed of 1 nm/s, with the laser power set to 0 dBm. An off-chip reference path normalizes the wavelength-dependent fiber-to-chip coupling response.

## Q-factor and propagation-loss extraction

Loaded $Q$ factors are derived from the Lorentzian 3-dB linewidth $\Delta f$ as $Q_{\mathrm{load}} = f_0/\Delta f$, where $f_0 = c/\lambda_0$ is the resonance frequency, $c$ the speed of light, and $\lambda_0$ the resonance wavelength. Intrinsic $Q$ factors are then obtained from $Q_{\mathrm{int}} = 2Q_{\mathrm{load}}/(1 \pm \sqrt{T_{\mathrm{min}}})$, where $T_{\mathrm{min}}$ is the on-resonance transmission minimum, with the (+) sign for under-coupling and the (−) for over-coupling. The two regimes are distinguished by sweeping the bus-to-ring coupling gap and tracking the direction of the $Q_{\mathrm{load}}$ variation. Propagation loss is calculated as $\alpha = 2\pi n_{\mathrm{g}}/(Q_{\mathrm{int}}\lambda) \cdot 4.343$, where $n_{\mathrm{g}}$ is the $\mathrm{TE_0}$ group index and the factor $4.343 = 10/\ln 10$ converts the loss from nepers to decibels, with $n_{\mathrm{g}} = 3.762$ ($W = 1.6$ μm) and 3.748 ($W = 2.0$ μm) from 2D eigenmode simulations at $\lambda = 1550$ nm.

---

## References

[1] Bogaerts, W. et al. Programmable photonic circuits. *Nature* **586**, 207–216 (2020).

[2] Sun, C. et al. Single-chip microprocessor that communicates directly using light. *Nature* **528**, 534–538 (2015).

[3] Rahim, A., Spuesens, T., Baets, R. & Bogaerts, W. Open-access silicon photonics: current status and emerging initiatives. *Proc. IEEE* **106**, 2313–2330 (2018).

[4] Daudlin, S. et al. Three-dimensional photonic integration for ultra-low-energy, high-bandwidth interchip data links. *Nat. Photonics* **19**, 502–509 (2025).

[5] Bandyopadhyay, S. et al. Single-chip photonic deep neural network with forward-only training. *Nat. Photonics* **18**, 1335–1343 (2024).

[6] Ahmed, S. R. et al. Universal photonic artificial intelligence acceleration. *Nature* **640**, 368–374 (2025).

[7] Marpaung, D., Yao, J. & Capmany, J. Integrated microwave photonics. *Nat. Photonics* **13**, 80–90 (2019).

[8] Daulay, O. et al. Ultrahigh dynamic range and low noise figure programmable integrated microwave photonic filter. *Nat. Commun.* **13**, 7798 (2022).

[9] Pérez-López, D. et al. General-purpose programmable photonic processor for advanced radiofrequency applications. *Nat. Commun.* **15**, 1563 (2024).

[10] Wang, J., Sciarrino, F., Laing, A. & Thompson, M. G. Integrated photonic quantum technologies. *Nat. Photonics* **14**, 273–284 (2020).

[11] Wang, J. et al. Multidimensional quantum entanglement with large-scale integrated optics. *Science* **360**, 285–291 (2018).

[12] Aghaee Rad, H. et al. Scaling and networking a modular photonic quantum computer. *Nature* **638**, 912–919 (2025).

[13] Lee, K. K., Lim, D. R., Kimerling, L. C., Shin, J. & Cerrina, F. Fabrication of ultralow-loss $Si/SiO_2$ waveguides by roughness reduction. *Opt. Lett.* **26**, 1888–1890 (2001).

[14] Zhang, L. et al. Ultralow-loss silicon photonics beyond the single-mode regime. *Laser Photonics Rev.* **16**, 2100292 (2022).

[15] Li, C., Liu, D. & Dai, D. Multimode silicon photonics. *Nanophotonics* **8**, 227–247 (2019).

[16] Wang, L. et al. Ultra-low loss and ultra-compact polarization-insensitive SOI multimode waveguide crossing based on an inverse design method. *Photonics* **11**, 1137 (2024).

[17] Sun, A. et al. Edge-guided inverse design of digital metamaterial-based mode multiplexers for high-capacity multi-dimensional optical interconnect. *Nat. Commun.* **16**, 2372 (2025).

[18] Qiu, H. et al. A continuously tunable sub-gigahertz microwave photonic bandpass filter based on an ultra-high-Q silicon microring resonator. *J. Lightwave Technol.* **36**, 4312–4318 (2018).

[19] Zhang, Y. et al. Design and demonstration of ultra-high-Q silicon microring resonator based on a multi-mode ridge waveguide. *Opt. Lett.* **43**, 1586–1589 (2018).

[20] Tao, Z. et al. Versatile photonic molecule switch in multimode microresonators. *Light Sci. Appl.* **13**, 51 (2024).

[21] Cherchi, M., Ylinen, S., Harjanne, M., Kapulainen, M. & Aalto, T. Dramatic size reduction of waveguide bends on a micron-scale silicon photonic platform. *Opt. Express* **21**, 17814–17823 (2013).

[22] Bahadori, M., Nikdast, M., Cheng, Q. & Bergman, K. Universal design of waveguide bends in silicon-on-insulator photonics platform. *J. Lightwave Technol.* **37**, 3044–3054 (2019).

[23] Vogelbacher, F. et al. Analysis of silicon nitride partial Euler waveguide bends. *Opt. Express* **27**, 31394–31406 (2019).

[24] Jiang, X., Wu, H. & Dai, D. Low-loss and low-crosstalk multimode waveguide bend on silicon. *Opt. Express* **26**, 17680–17689 (2018).

[25] Kita, S. et al. Wide waveguide concatenated Euler bends for compact and fabrication error-tolerant pseudo-single mode silicon photonics. In *Proc. Conference on Lasers and Electro-Optics (CLEO)* paper SF3F.3 (Optica Publishing Group, 2024).

[26] Sun, S. et al. Inverse design of ultra-compact multimode waveguide bends based on the free-form curves. *Laser Photonics Rev.* **15**, 2100162 (2021).

[27] Sun, C., Yu, Y., Chen, G. & Zhang, X. Ultra-compact bent multimode silicon waveguide with ultralow inter-mode crosstalk. *Opt. Lett.* **42**, 3004–3007 (2017).

[28] Liu, Y. et al. Very sharp adiabatic bends based on an inverse design. *Opt. Lett.* **43**, 2482–2485 (2018).

[29] Irfan, S., Kim, J. Y. & Kurt, H. Ultra-compact and efficient photonic waveguide bends with different configurations designed by topology optimization. *Sci. Rep.* **14**, 6453 (2024).

[30] Han, Z. et al. Design of ultrahigh-Q silicon microring resonators based on free-form curves. *Opt. Express* **32**, 9553–9561 (2024).

[31] Zhang, L. et al. Ultrahigh-Q silicon racetrack resonators. *Photonics Res.* **8**, 684–689 (2020).

[32] Liu, Y., Choudhary, A., Marpaung, D. & Eggleton, B. J. Integrated microwave photonic filters. *Adv. Opt. Photonics* **12**, 485–555 (2020).

[33] Yao, J. & Capmany, J. Microwave photonics. *Sci. China Inf. Sci.* **65**, 221401 (2022).

[34] Marpaung, D. et al. $Si_3N_4$ ring resonator-based microwave photonic notch filter with an ultrahigh peak rejection. *Opt. Express* **21**, 23286–23294 (2013).

[35] Zeng, D. et al. Demonstration of ultra-high-Q silicon microring resonators for nonlinear integrated photonics. *Micromachines* **13**, 1155 (2022).

[36] Nijem, J. et al. High-Q and high finesse silicon microring resonator. *Opt. Express* **32**, 7896–7906 (2024).

[37] Guillén-Torres, M. Á. et al. Effects of backscattering in high-Q, large-area silicon-on-insulator ring resonators. *Opt. Lett.* **41**, 1538–1541 (2016).

[38] Wang, C. et al. Integrated lithium niobate electro-optic modulators operating at CMOS-compatible voltages. *Nature* **562**, 101–104 (2018).

[39] Boes, A., Corcoran, B., Chang, L., Bowers, J. & Mitchell, A. Status and potential of lithium niobate on insulator (LNOI) for photonic integrated circuits. *Laser Photonics Rev.* **12**, 1700256 (2018).

[40] Wei, C. et al. Programmable multifunctional integrated microwave photonic circuit on thin-film lithium niobate. *Nat. Commun.* **16**, 2281 (2025).

[41] Fandiño, J. S., Muñoz, P., Domènech, D. & Capmany, J. A monolithic integrated photonic microwave filter. *Nat. Photonics* **11**, 124–129 (2017).

[42] Bauters, J. F. et al. Ultra-low-loss high-aspect-ratio $Si_3N_4$ waveguides. *Opt. Express* **19**, 3163–3174 (2011).

[43] Pfeiffer, M. H. P. et al. Photonic damascene process for low-loss, high-confinement silicon nitride waveguides. *IEEE J. Sel. Top. Quantum Electron.* **24**, 6101411 (2018).

[44] Ye, Z. et al. Foundry manufacturing of tight-confinement, dispersion-engineered, ultralow-loss silicon nitride photonic integrated circuits. *Photonics Res.* **11**, 558–568 (2023).

[45] Bose, D. et al. Anneal-free ultra-low loss silicon nitride integrated photonics. *Light Sci. Appl.* **13**, 156 (2024).

# Supplementary Information for Smooth-Curvature Bend Design Guided by Variational Analysis for Adiabatic Multimode Integrated Photonics

**Yutong Shi[1,2,*], Shijin Yu[1,2,*], Ye Huang[1,2], Yuan Yu[1,2,3,✉], and Xinliang Zhang[1,2,3,✉]**

[1]Wuhan National Laboratory for Optoelectronics, Huazhong University of Science and Technology, Wuhan 430074, China

[2]School of Optical and Electronic Information, Huazhong University of Science and Technology, Wuhan 430074, China

[3]Optics Valley Laboratory, Wuhan 430074, China

✉ e-mail: yuan_yu@hust.edu.cn; xlzhang@mail.hust.edu.cn

---

## S1. Sidewall scattering model and the multimode-width advantage

The first design pillar of the smooth polynomial curvature (SPC) framework, a wide multimode core, is motivated by the width dependence of sidewall scattering loss. Fig. S1a presents the simulated fundamental transverse-electric ($TE_0$) mode $|E|^2$ profiles of a 450 nm single-mode core and a 1.6 μm multimode core on the 220 nm silicon-on-insulator (SOI) strip waveguide. Although the field peak sits at the core center in both cases, the narrow single-mode core retains appreciable field intensity at the etched sidewalls, whereas the wide multimode core confines the field well within the core interior. To quantify the resulting impact on propagation loss, we adopted the perturbative sidewall scattering model (the "nw" model) of Melati et al. [S1], treating the etched sidewall as a one-dimensional random profile with root-mean-square (RMS) height $\sigma$ and correlation length $L_c$. The $TE_0$ scattering loss coefficient is given by

$$\alpha_{sc}(W) = \frac{2\sigma^2}{k_0\, n_{core}^2} \left| \frac{\partial n_{eff}}{\partial W} \right|^2 F(\beta, L_c), \tag{S1}$$

where $k_0 = 2\pi/\lambda$ is the vacuum wavenumber, $\lambda$ is the operating wavelength, $n_{core}$ is the core refractive index, $W$ is the waveguide core width, $n_{eff}$ is the $TE_0$ effective index, $\partial n_{eff}/\partial W$ is the sensitivity of the $TE_0$ effective index to the core width, $\beta$ is the propagation constant, and $F$ is a geometry factor determined by the surface autocorrelation function.

We compute $n_{\mathrm{eff}}(W)$ for a 220 nm SOI strip waveguide with $n_{\mathrm{core}}$ = 3.476 (Si) and $n_{\mathrm{clad}}$ = 1.444 ($SiO_2$) at $\lambda$ = 1550 nm using a 2D finite-difference eigenmode solver (Lumerical MODE), from which $\partial n_{\mathrm{eff}}/\partial W$ is extracted. The correlation length is set to $L_c$ = 50 nm, a value typical of 193 nm deep-ultraviolet lithography on standard SOI.

Fig. S1b shows the effective-index sensitivity to width $|\partial n_{\mathrm{eff}}/\partial W|$ as a function of $W$. The sensitivity decreases monotonically with increasing $W$, from 1.97 μm⁻¹ at $W$ = 0.45 μm to 0.026 μm⁻¹ at $W$ = 2.0 μm, a reduction of nearly two orders of magnitude. Fig. S1c shows $\alpha_{\mathrm{sc}}(W)$ for three RMS roughness levels ($\sigma$ = 2, 4, and 7 nm). The three curves are scaled replicas of one another, related through the $\sigma^2$ prefactor in Equation (S1), and all decrease monotonically with $W$, mirroring the width dependence of $|\partial n_{\mathrm{eff}}/\partial W|^2$.

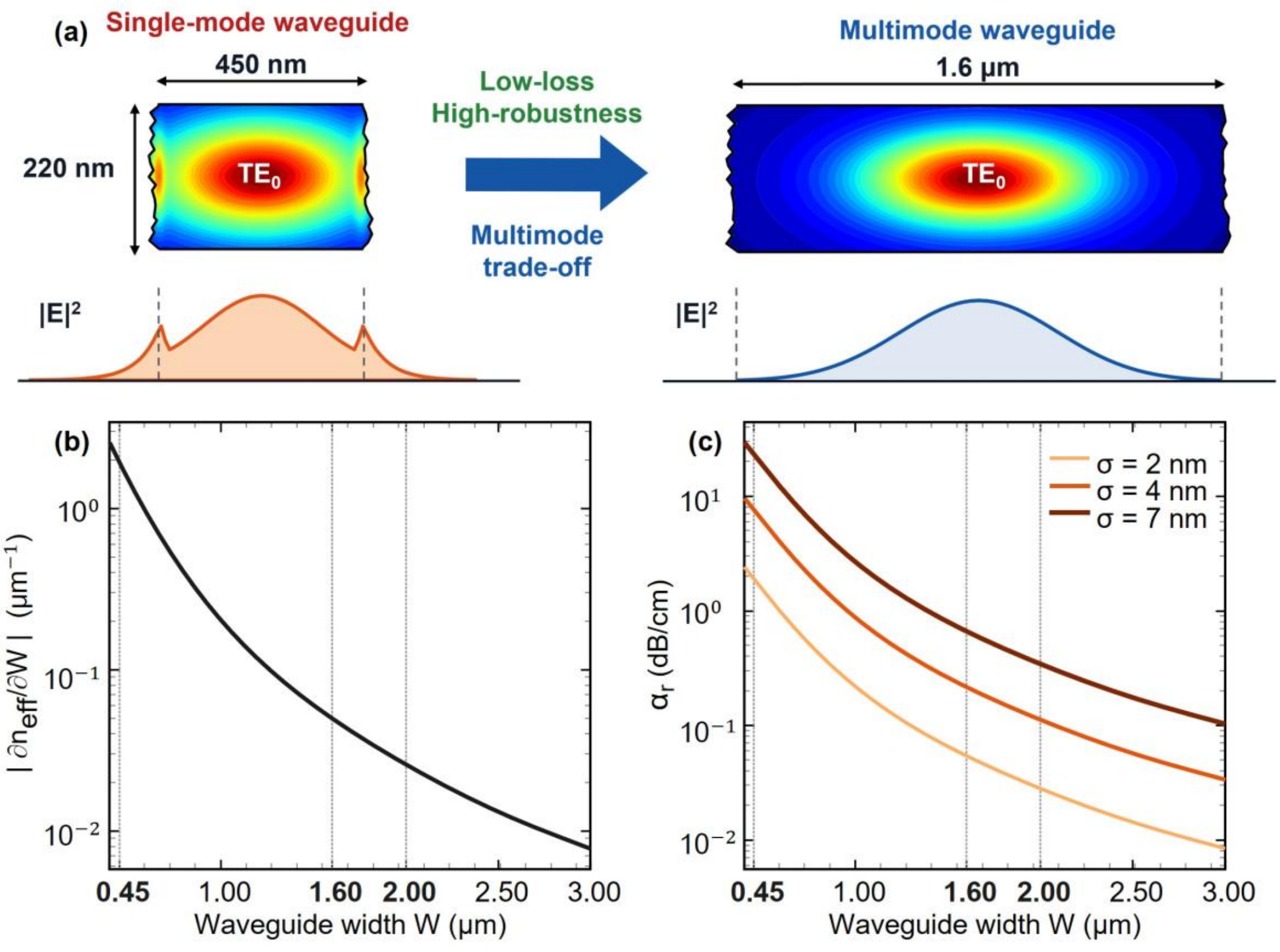

**Figure S1.** Width dependence of mode confinement and sidewall-scattering loss in the multimode SOI strip waveguide. **a**, Single-mode ($W$ = 450 nm) and multimode ($W$ = 1.6 μm) 220 nm SOI strip waveguides with simulated $TE_0$ $|E|^2$ profiles. The narrow single-mode core retains appreciable field intensity at the etched sidewalls, whereas the wide multimode core confines the field to the core interior. **b**, $TE_0$ effective-index sensitivity to width, $|\partial n_{\mathrm{eff}}/\partial W|$, for the 220 nm SOI strip waveguide at $\lambda$ = 1550 nm. **c**, Sidewall scattering loss $\alpha_{\mathrm{sc}}$ for sidewall RMS roughness $\sigma$ = 2, 4, and 7 nm with correlation length $L_c$ = 50 nm. Vertical dashed lines in **b**, **c** mark the single-mode reference width (0.45 μm) and the two main-text multimode widths (1.6 and 2.0 μm).

This monotonic decrease confirms that wider multimode cores suppress sidewall scattering loss, justifying our choice of $W$ = 1.6 and 2.0 μm for the smooth polynomial curvature hybrid (SPCh) microring resonator (MRR) devices presented in the main text.

---

## S2. Eigenmode analysis of multimode-waveguide bends

To establish the quantitative relationship between curvature and modal behavior, we simulate multimode strip waveguides on the 220 nm silicon-on-insulator (SOI) platform with a 1.6 μm core width and $SiO_2$ cladding. Eigenmodes are obtained using a 2D finite-difference eigenmode solver (Lumerical MODE). By evaluating overlap integrals between eigenmodes at different local curvatures, we obtain three metrics. The higher-order mode (HOM) excitation ratio and the fundamental-mode mismatch loss depend on the curvature difference $\Delta\kappa$ between adjacent sections, whereas the normalized sidewall field intensity depends on the local curvature $\kappa$. We fit these simulated dependences with low-order functions for use in the variational analysis.

The eigenmode simulations show that the $TE_0$ field shifts toward the outer sidewall as $\kappa$ increases. This curvature-induced change in the local mode profile alters mode coupling between adjacent sections and increases the sidewall field intensity, providing the physical basis for the three metrics discussed below.

### S2.1 Higher-order mode excitation

The HOM excitation ratio $\eta_{\mathrm{HOM}}$ induced by a curvature difference $\Delta\kappa$ between adjacent sections is fitted with a quartic function of $\Delta\kappa$, as plotted in Fig. 2c of the main text:

$$\eta_{\mathrm{HOM}}(\Delta\kappa) \approx \alpha_1 \Delta\kappa + \alpha_2 (\Delta\kappa)^2 + \alpha_3 (\Delta\kappa)^3 + \alpha_4 (\Delta\kappa)^4. \tag{S2}$$

The coefficients $\alpha_1$–$\alpha_4$ are obtained by a least-squares fit to the simulated HOM excitation ratio as a function of $\Delta\kappa$. The quartic provides a compact empirical representation of the simulated dependence. This fitted functional form, rather than the individual coefficient values, is used to construct the figure-of-merit functional and support the variational analysis.

### S2.2 Fundamental-mode mismatch loss

The $TE_0$ mismatch loss $L_{\mathrm{mis}}$ between sections of different curvature is likewise fitted with a quartic function of $\Delta\kappa$, as plotted in Fig. 2d of the main text:

$$L_{\mathrm{mis}}(\Delta\kappa) \approx \beta_1 \Delta\kappa + \beta_2 (\Delta\kappa)^2 + \beta_3 (\Delta\kappa)^3 + \beta_4 (\Delta\kappa)^4, \tag{S3}$$

The coefficients $\beta_1$–$\beta_4$ are obtained by a least-squares fit to the simulated $L_{\text{mis}}$ as a function of $\Delta\kappa$.

Equations (S2) and (S3) describe two outputs of the same local mode-coupling calculation for adjacent sections with curvature difference $\Delta\kappa$. The HOM excitation ratio describes the power coupled to HOMs, whereas $L_{\text{mis}} = -10\log_{10}T_{\text{TE}_0}$ describes the reduction in $\text{TE}_0$ transmission. The two metrics are fitted independently, and the quartic functions provide smooth empirical representations of the simulated data for the variational analysis.

## S2.3 Sidewall field intensity

The optical field intensity at the waveguide sidewall, which is an input to the sidewall scattering-loss model, is fitted with a cubic function of the absolute curvature $\kappa$, as plotted in Fig. 2e of the main text:

$$\rho_{\text{sw}}(\kappa) \approx \gamma_0 + \gamma_1\kappa + \gamma_2\kappa^2 + \gamma_3\kappa^3, \tag{S4}$$

The coefficients $\gamma_0$–$\gamma_3$ are obtained by a least-squares fit to the simulated $\rho_{\text{sw}}(\kappa)$, where $\rho_{\text{sw}}$ is the sidewall field intensity normalized to its straight-waveguide value so that the $\gamma$ coefficients are dimensionless. As $\kappa$ increases, the $\text{TE}_0$ mode is progressively pushed toward the outer sidewall. The constant term $\gamma_0$ represents the straight-waveguide baseline at $\kappa = 0$. The linear and quadratic terms describe this perturbative outward shift of the mode field. The cubic term accounts for the additional field accumulation at the outer sidewall that becomes significant at stronger curvatures.

## S2.4 Construction of the figure-of-merit functional

Equations (S2) and (S3) describe the two modal metrics as functions of the discrete curvature difference $\Delta\kappa$ between adjacent waveguide sections, whereas Equation (S4) describes the sidewall metric as a function of the local curvature $\kappa$. To use these fitted functions in the continuous variational framework, consider adjacent sections separated by an arc-length interval $\delta s$. Their curvature difference satisfies $\Delta\kappa = \kappa(s+\delta s) - \kappa(s) \approx \kappa'\,\delta s$ to leading order, where $\kappa' = d\kappa/ds$. Substituting this relation maps each $\Delta\kappa$-dependent term in Equations (S2) and (S3) to a term in $\kappa'$, with each factor $(\delta s)^k$ absorbed into the corresponding renormalized coefficient. Equation (S4) directly contributes terms in the local curvature $\kappa$.

After this discrete-to-continuous mapping, the two modal metrics contribute terms in $\kappa'$, while the sidewall metric contributes terms in $\kappa$. To combine them into a single quantitative metric of multimode waveguide bend (MWB) performance that is suitable for variational optimization, we construct a polynomial Lagrangian density and define the figure-of-merit (FOM) functional as its integral along the bend. The coefficients $\alpha_1$–$\alpha_4$ in Equation (S2) (HOM excitation ratio) and $\beta_1$–$\beta_4$ in Equation (S3) (fundamental-mode mismatch loss) renormalize into four dimensionless weights $E$, $F$, $G$, $H$ that scale $\kappa'$, $(\kappa')^2$, $(\kappa')^3$, and $(\kappa')^4$ in the Lagrangian density. The sidewall coefficients $\gamma_0$–$\gamma_3$ in Equation (S4) (sidewall optical field intensity) renormalize into $A$, $B$, $C$, $D$ that scale the constant, $\kappa$, $\kappa^2$, and $\kappa^3$ terms. The resulting eight coefficients $A-H$ thus serve as a single set of weighted normalization factors that fold the three metrics into the polynomial Lagrangian density of the FOM functional:

$$\mathrm{FOM}[\kappa] = \int_{s_1}^{s_2} \left[A + B\kappa + C\kappa^2 + D\kappa^3 + E\kappa' + F\kappa'^2 + G\kappa'^3 + H\kappa'^4\right] ds. \tag{S5}$$

Here $s$ is the normalized arc-length coordinate of the complete bend, with $s \in [0,2]$, $s_1 = 0$ at the bend entrance, $s = 1$ at the midpoint, and $s_2 = 2$ at the bend exit. The curvature $\kappa$ is scaled by $\kappa_{\max} = 1/R_{\min}$, with $R_{\min}$ the minimum bend radius, so that $A$–$H$ are dimensionless real constants whose values depend only on the platform and the operating wavelength, not on the bend radius. Minimizing this FOM seeks a curvature profile that reduces the weighted contributions from HOM excitation, $TE_0$ mismatch loss, and sidewall field intensity. The polynomial structure of the integrand underlies the infinite-differentiability necessary condition stated in the main text. The proof in Section S3 depends solely on this polynomial form and not on the specific values of $A$–$H$.

---

## S3. Variational derivation of the infinite-differentiability necessary condition

This section establishes the universal infinite-differentiability regularity result stated in the main text. The proof proceeds in three steps. First, we derive the Euler–Lagrange optimality ordinary differential equation (ODE) from the FOM functional [Equation (S5)]. Second, we show that its right-hand side is infinitely differentiable under the Legendre necessary condition. Third, we recursively lift the minimizer's regularity from twice continuously differentiable to infinitely differentiable.

### S3.1 Euler–Lagrange optimality equation

Following the standard framework of variational calculus, we identify the integrand of Equation (S5) as the Lagrangian density $\mathscr{L}(\kappa, \kappa')$. The Euler–Lagrange equation requires

$$\frac{\partial \mathscr{L}}{\partial \kappa} - \frac{d}{ds}\frac{\partial \mathscr{L}}{\partial \kappa'} = 0. \tag{S6}$$

Computing each term yields

$$\frac{\partial \mathscr{L}}{\partial \kappa} = B + 2C\kappa + 3D\kappa^2, \tag{S7a}$$

and

$$\frac{\partial \mathcal{L}}{\partial \kappa'} = E + 2F\kappa' + 3G\kappa'^2 + 4H\kappa'^3. \tag{S7b}$$

Since $\partial \mathcal{L}/\partial \kappa'$ depends solely on $\kappa'$, its total derivative with respect to arc length $s$ reads

$$\frac{d}{ds}\frac{\partial \mathcal{L}}{\partial \kappa'} = \frac{\partial^2 \mathcal{L}}{\partial \kappa'^2}\kappa'' = (2F + 6G\kappa' + 12H\kappa'^2)\kappa''. \tag{S7c}$$

Substituting Equations (S7a) and (S7c) into Equation (S6) and rearranging yields the second-order nonlinear ODE governing the optimal curvature:

$$\kappa''(s) = \frac{B+2C\kappa+3D\kappa^2}{2F+6G\kappa'+12H\kappa'^2}. \tag{S8}$$

This is the general Euler–Lagrange optimality equation, derived without approximations or coefficient simplifications.

## S3.2 Legendre necessary condition for local minimizers

The Euler–Lagrange equation identifies stationary points of the FOM functional, including minima, maxima, and saddle points. To isolate local minimizers, the Legendre necessary condition requires:

$$\frac{\partial^2 \mathcal{L}}{\partial \kappa'^2} \geq 0 \tag{S9}$$

evaluated along the minimizing $\kappa(s)$.

Computing the second partial derivative gives

$$\frac{\partial^2 \mathcal{L}}{\partial \kappa'^2} = 2F + 6G\kappa' + 12H\kappa'^2 \equiv P(\kappa'). \tag{S10}$$

Since the global minimizer is necessarily a local minimizer, the same condition applies to the global optimum. We further establish that $P(\kappa') > 0$ holds strictly and globally. $P(\kappa')$ is a quadratic polynomial in $\kappa'$, with leading coefficient $12H$ and discriminant $36G^2 - 96FH$. Strict positivity for all real $\kappa'$ is therefore equivalent to two standard algebraic conditions: a positive leading coefficient (so that the parabola opens upward) and a negative discriminant (so that the parabola has no real roots). The leading coefficient $12H$ requires $H > 0$, since a non-positive $H$ would leave $\mathcal{L}$ unbounded below at large $|\kappa'|$, driving the FOM arbitrarily negative. The discriminant condition $\Delta_D = 36G^2 - 96FH < 0$ rules out any real root of $P(\kappa')$, since an extremal passing through such a root would violate the strengthened Legendre condition.

With both inequalities satisfied, the denominator $P(\kappa')$ in Equation (S8) is strictly positive for all real $\kappa'$. Defining the right-hand side of Equation (S8) as $\Phi(\kappa, \kappa') = N(\kappa)/P(\kappa')$, the numerator $N(\kappa) = B + 2C\kappa + 3D\kappa^2$ is a polynomial and thus infinitely differentiable over $\mathbb{R}$. Consequently, $\Phi(\kappa, \kappa')$ is infinitely differentiable over the entire $\mathbb{R}^2$ plane.

## S3.3 Recursive lift from twice differentiable to infinitely differentiable

Let $\kappa(s)$ be a minimizing solution on the bend interval $I = [s_1, s_2]$. By definition, $\kappa(s)$ is at least twice continuously differentiable, so $\kappa'$ and $\kappa''$ are continuous on $I$. The Euler–Lagrange equation $\kappa''(s) = \Phi(\kappa(s), \kappa'(s))$ expresses $\kappa''$ as an infinitely differentiable function of continuous arguments. Differentiating both sides once gives

$$\kappa'''(s) = \frac{\partial \Phi}{\partial \kappa}\,\kappa' + \frac{\partial \Phi}{\partial \kappa'}\,\kappa''. \tag{S11}$$

Since the partial derivatives of $\Phi$ are continuous and both $\kappa'$ and $\kappa''$ are continuous, $\kappa'''(s)$ exists and is continuous, so $\kappa$ is three-times continuously differentiable. Through recursion, every higher-order derivative $\kappa^{(n)}$ is a continuous combination of lower-order derivatives and the smooth function $\Phi$ together with its partial derivatives. The regularity of $\kappa$ therefore extends to all orders, and the minimizing curvature is infinitely differentiable on the entire interval $I$.

Therefore, infinite smoothness is an intrinsic property and a necessary condition for $\kappa(s)$ to satisfy the Euler–Lagrange equation while minimizing the FOM functional. Geometrically, the optimal curvature must be continuous in $\kappa$ and in all its derivatives. This regularity is necessary for minimizing the modeled FOM, but it is not by itself sufficient to determine the optimal finite-order bend profile.

## S3.4 Physical implications

Infinite differentiability is a necessary condition for $\kappa(s)$ to simultaneously satisfy the Euler–Lagrange equation and minimize the FOM functional. The proof depends only on the polynomial structure of the Lagrangian and on the Legendre condition. No specific coefficient values or platform parameters are required. The conclusion is therefore universal: regardless of waveguide geometry, material, or wavelength, the optimal curvature must be infinitely differentiable.

This necessary condition rules out both circular arcs and Euler bends as optimal solutions of the FOM functional. Circular arcs exhibit a discontinuous $\kappa$, while Euler bends exhibit a discontinuous $\kappa'$. The Euler bend represents a progressive improvement over the circular arc, gaining continuity of $\kappa$ but still failing continuity of $\kappa'$. Both designs nevertheless remain suboptimal in the variational sense established by Equations (S8) and (S10).

This necessary condition shows that any FOM minimizer is infinitely differentiable, not that finite-order designs are excluded. A finite-order polynomial satisfying the junction conditions closely approaches the smoothness of an infinitely differentiable profile. The best order for a given bend therefore depends on its minimum radius, not on smoothness alone.

---

# S4. SPC curvature polynomials and the SPC hybrid

This section derives the SPC curvature family and motivates the choice of the SPC hybrid (SPCh) employed throughout the main text.

## S4.1 SPC curvature polynomial and boundary conditions

The infinite-differentiability requirement established in the main text is directly realized via an odd-order polynomial curvature profile defined over the first half of the normalized arc length, $s \in [0, 1]$, and mirrored about the midpoint to form the full symmetric bend (Section S8):

$$\kappa_n(s) = \kappa_{\max} \sum_{k=0}^{n} a_k \, s^k, \tag{S12}$$

where $\kappa_{\max} = 1/R_{\min}$ is the maximum curvature at the bend midpoint, $a_k$ are the polynomial coefficients fixed by the junction constraints below, $n = 2m + 1$ is the polynomial order, and $m \in \mathbb{N}^+$.

This choice ensures the number of polynomial coefficients ($n + 1$) exactly matches the $2(m + 1)$ paired boundary conditions required for $m$-th-order curvature continuity, yielding a well-determined, full-rank linear system. The junction values are pinned by $\kappa(0) = 0$ and $\kappa(1) = \kappa_{\max}$, while the higher-order derivatives satisfy

$$\kappa^{(j)}(0) = \kappa^{(j)}(1) = 0, \quad j = 1,2, \dots, m. \tag{S13}$$

The $n$-th-order SPC bend therefore achieves an optimal transition from the straight waveguide ($\kappa = 0$) to the maximum curvature, with all derivatives up to order $m$ vanishing at both junctions.

## S4.2 Explicit SPC curvature functions

Solving the boundary conditions in Equation (S13) yields an explicit polynomial for each value of $m$. The three orders employed in the main text are listed below.

**3rd-order SPC ($n = 3, m = 1$).** The bend ensures continuity of $\kappa$ and its first derivative $\kappa'$ (both vanishing at the junctions):

$$\kappa_3(s) = \kappa_{\max} \, (3s^2 - 2s^3). \tag{S14}$$

**5th-order SPC ($n = 5, m = 2$).** The bend ensures continuity of $\kappa$ and its first two derivatives $\kappa'$ and $\kappa''$ (all vanishing at the junctions):

$$\kappa_5(s) = \kappa_{\max}\,(10s^3 - 15s^4 + 6s^5). \tag{S15}$$

**7th-order SPC ($n = 7, m = 3$).** The bend ensures continuity of $\kappa$ and its first three derivatives (all vanishing at the junctions):

$$\kappa_7(s) = \kappa_{\max}\,(35s^4 - 84s^5 + 70s^6 - 20s^7). \tag{S16}$$

Each SPC profile is fully determined by $\kappa_{\max}$ and the polynomial order $n$, with no free fitting parameters. The five-design comparison in Fig. 3a of the main text is constructed from Equations (S14)–(S16). The general term also encompasses the conventional Euler bend as its lowest-order $n = 1$ ($m = 0$) case, for which $\kappa_1(s) = \kappa_{\max}\,s$ varies linearly along the arc.

## S4.3 Unified Beta-function form

Analyzing the first derivative $\kappa_n'(s)$ of the boundary-condition-satisfying curvature function $\kappa_n(s)$ reveals a unified factored form:

$$\kappa_n'(s) = C_m\,s^m(1-s)^m, \qquad m = (n-1)/2, \tag{S17}$$

where $C_m$ is a normalization constant. The factors $s^m$ and $(1-s)^m$ intrinsically guarantee that $\kappa_n'(s)$ and all its derivatives up to order $m - 1$ vanish at $s = 0$ and $s = 1$, so that the boundary conditions in Equation (S13) are satisfied automatically. The normalization condition $\kappa_n(1) = \int_0^1 \kappa_n'(s)\,ds = \kappa_{\max}$ then fixes $C_m$:

$$C_m = \frac{\kappa_{\max}}{\int_0^1 s^m(1-s)^m\,ds} = \kappa_{\max}\,\frac{\Gamma(2m+2)}{[\Gamma(m+1)]^2} = \kappa_{\max}\,\frac{(2m+1)!}{(m!)^2}, \tag{S18}$$

where $\Gamma$ denotes the Gamma function and the second equality uses the standard identity $\int_0^1 s^m\,(1-s)^m\,ds = [\Gamma(m+1)]^2/\Gamma(2m+2) = B(m+1, m+1)$, with $B(p,q)$ the complete Beta function. Integrating Equation (S17) from 0 to $s$ then yields the unified curvature function

$$\kappa_n(s) = \kappa_{\max}\,\frac{\int_0^s t^m(1-t)^m\,dt}{\int_0^1 t^m(1-t)^m\,dt} = \kappa_{\max}\,\frac{B(s; m+1, m+1)}{B(m+1, m+1)} = \kappa_{\max}\,I_s(m+1, m+1), \tag{S19}$$

where $B(s;\,p,\,q) = \int_0^s t^{p-1}\,(1-t)^{q-1}\,dt$ is the incomplete Beta function and $I_s(p,q) = B(s;\,p,\,q)/B(p,\,q)$ is its regularized form. This compact form gives the general term of the SPC family for arbitrary odd order $n = 2m + 1$, with Equations (S14)–(S16) recovering $m$ = 1, 2, and 3. In this form, the five designs compared in Fig. 3a of the main text are $\kappa_3 = \kappa_{\max}\,I_s(2{,}2)$, $\kappa_5 = \kappa_{\max}\,I_s(3{,}3)$, $\kappa_7 = \kappa_{\max}\,I_s(4{,}4)$, the hybrid $\kappa_{\mathrm{SPCh}} = \kappa_{\max}\,[0.8\,I_s(2{,}2) + 0.2\,I_s(4{,}4)]$ introduced in Section S4.4, and $\kappa_{\mathrm{Euler}} = \kappa_{\max}\,I_s(1{,}1) = \kappa_{\max}\,s$.

Expanding $(1-t)^m$ in Equation (S19) gives the explicit polynomial-coefficient form

$$\frac{\kappa_{2m+1}(s)}{\kappa_{\max}} = \frac{(2m+1)!}{(m!)^2}\sum_{j=0}^{m}\frac{(-1)^j\binom{m}{j}}{m+j+1}s^{m+j+1}. \tag{S20}$$

Equation (S20) directly generates the coefficients of any odd-order SPC profile from $m$. For example, setting $m$ = 4 gives the ninth-order profile

$$\kappa_9(s) = \kappa_{\max}\,(126s^5 - 420s^6 + 540s^7 - 315s^8 + 70s^9). \tag{S21}$$

## S4.4 Hybrid SPC design

While higher-order SPC bends satisfy more boundary conditions on curvature continuity, they are not universally superior in practice. As the polynomial order n increases, the curvature must traverse its full range within the same arc length while meeting stricter junction constraints. This forces the curvature gradient $\kappa'$ to concentrate in the central region of the bend: the junction smoothness improves, but $\kappa'$ becomes more sharply peaked at the center. Fig. 2i of the main text illustrates this trend across SPC-3, SPC-5, SPC-7, and the Euler bend: as n grows from 3 to 7, the $\kappa'$ profile becomes increasingly peaked in the central region. Consequently, for a given minimum radius $R_{\min}$, there exists an optimal SPC order that reconciles junction smoothness with interior gradient.

To achieve this balance, we introduce a hybrid SPC profile as a weighted combination of two orders:

$$\kappa_{\text{hybrid}}(s) = \sum_i w_i\,\kappa_{n_i}(s), \qquad \sum_i w_i = 1. \tag{S22}$$

The hybrid inherits the smoothness contributions of its constituent orders while mitigating their individual drawbacks. We restrict the family to a two-term combination of SPC-3 and SPC-7,

$$\kappa_{\text{hybrid}}(s) = w_3\,\kappa_3(s) + w_7\,\kappa_7(s), \tag{S23}$$

with the weights satisfying $w_3 + w_7 = 1$.

We swept the weight parameter $w_3$ (with $w_7 = 1 - w_3$) at the compact-radius operating points of the main text on 220 nm SOI and identified $w_3 = 0.8$, $w_7 = 0.2$ as the joint optimum on the insertion-loss versus mode-extinction-ratio plane (Fig. 3a of the main text). The 80% SPC-3 contribution provides the moderate central $\kappa'$ gradient characteristic of low-order profiles, and the 20% SPC-7 contribution suppresses the residual junction discontinuity of the cubic SPC-3. As a result, SPCh exhibits a boundary smoothness intermediate between SPC-3 and SPC-7 while avoiding the central $\kappa'$ peak associated with pure SPC-7.

At a fixed minimum radius, the infinite-differentiability condition established in the main text is necessary but not sufficient for optimality. The variational analysis fixes the regularity of the minimizer but not a finite-order realization. However, increasing $m$ concentrates the curvature gradient into a narrower central region, producing steeper $\kappa'$ peaks that counteract the benefit of smoother junctions. The optimal performance is therefore achieved at a moderate order that balances these two competing effects, rather than by maximizing $m$. The gap between necessity and sufficiency is set by the $R_{\min}$ constraint: relaxing it would lengthen the available arc, reduce the central $\kappa'$ peak, and progressively favor smoother high-order profiles. SPCh represents this practical operating point at compact $R_{\min}$, balancing junction smoothness against the interior curvature gradient. The 80:20 weighting is locally optimal for the 220 nm SOI strip-waveguide platform. Repeating the weight sweep on a different platform, for example lithium niobate on insulator (LNOI) in Section S7, would in principle shift this optimum. Nevertheless, applying the same SPCh = 0.8 SPC-3 + 0.2 SPC-7 operating point on LNOI (Section S7) confirms that SPCh remains near-optimal, consistent with the platform-independent nature of the design principle.

---

## S5. Design-principle validation on 90° bends

The transmission spectra of 90° SPCh and Euler bends across compact and relaxed effective bend radii $R_{\mathrm{eff}}$ (twice the minimum radius, $R_{\mathrm{eff}} = 2R_{\min}$) are reported in Figs. S2 and S3. The compact configuration uses $R_{\min}$ = 10 μm at $W$ = 1.6 μm (effective radius 20 μm) and $R_{\min}$ = 12 μm at $W$ = 2.0 μm (effective radius 24 μm), respectively. The relaxed configuration uses $R_{\min}$ = 18 μm at $W$ = 1.6 μm (effective radius 36 μm) and $R_{\min}$ = 30 μm at $W$ = 2.0 μm (effective radius 60 μm), respectively.

At the compact radii shown in Fig. S2, SPCh attains a maximum mode extinction ratio ($\mathrm{MER}_{\max}$) of −27.1 dB at $W$ = 1.6 μm and −31.0 dB at $W$ = 2.0 μm across 1500–1600 nm. The Euler bend at the same $R_{\min}$ reaches $\mathrm{MER}_{\max}$ values of only −11.9 dB and −9.4 dB, respectively, corresponding to 15.1 dB and 21.7 dB lower mode suppression than SPCh. The maximum insertion loss ($\mathrm{IL}_{\max}$) remains below 0.012 dB for SPCh at both widths, whereas the Euler bend reaches 0.274 dB at $W$ = 1.6 μm and 0.487 dB at $W$ = 2.0 μm.

At the relaxed radii shown in Fig. S3, SPCh achieves $\mathrm{MER}_{\max}$ = −36.8 dB at $W$ = 1.6 μm and −38.0 dB at $W$ = 2.0 μm, with $\mathrm{IL}_{\max}$ below 0.004 dB at both widths. In contrast, the Euler bend at the same $R_{\min}$ reaches $\mathrm{MER}_{\max}$ values of only −22.3 dB and −20.3 dB, with $\mathrm{IL}_{\max}$ ranging from 0.027 to 0.043 dB. The SPCh-over-Euler MER advantage therefore remains substantial at the relaxed radii, reaching 14.5 dB and 17.8 dB.

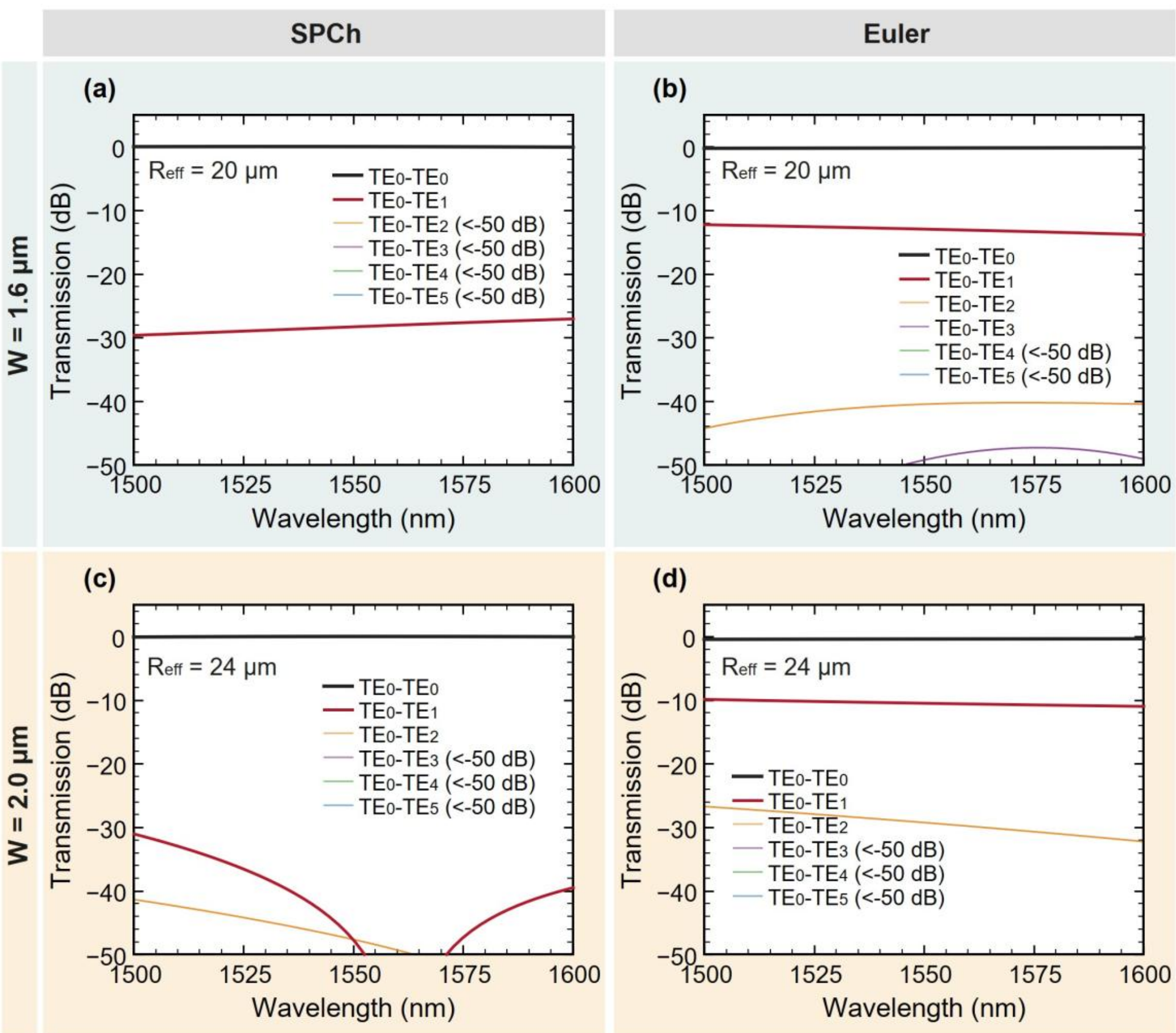

**Figure S2.** Wideband response of compact 90° bends. Transmission spectra of 90° bends with $TE_0$ launched and the output decomposed onto $TE_0$–$TE_5$ across 1500–1600 nm. **a**, SPCh bend at $W$ = 1.6 μm and $R_{\text{eff}}$ = 20 μm. **b**, Euler bend at $W$ = 1.6 μm and $R_{\text{eff}}$ = 20 μm. **c**, SPCh bend at $W$ = 2.0 μm and $R_{\text{eff}}$ = 24 μm. **d**, Euler bend at $W$ = 2.0 μm and $R_{\text{eff}}$ = 24 μm.

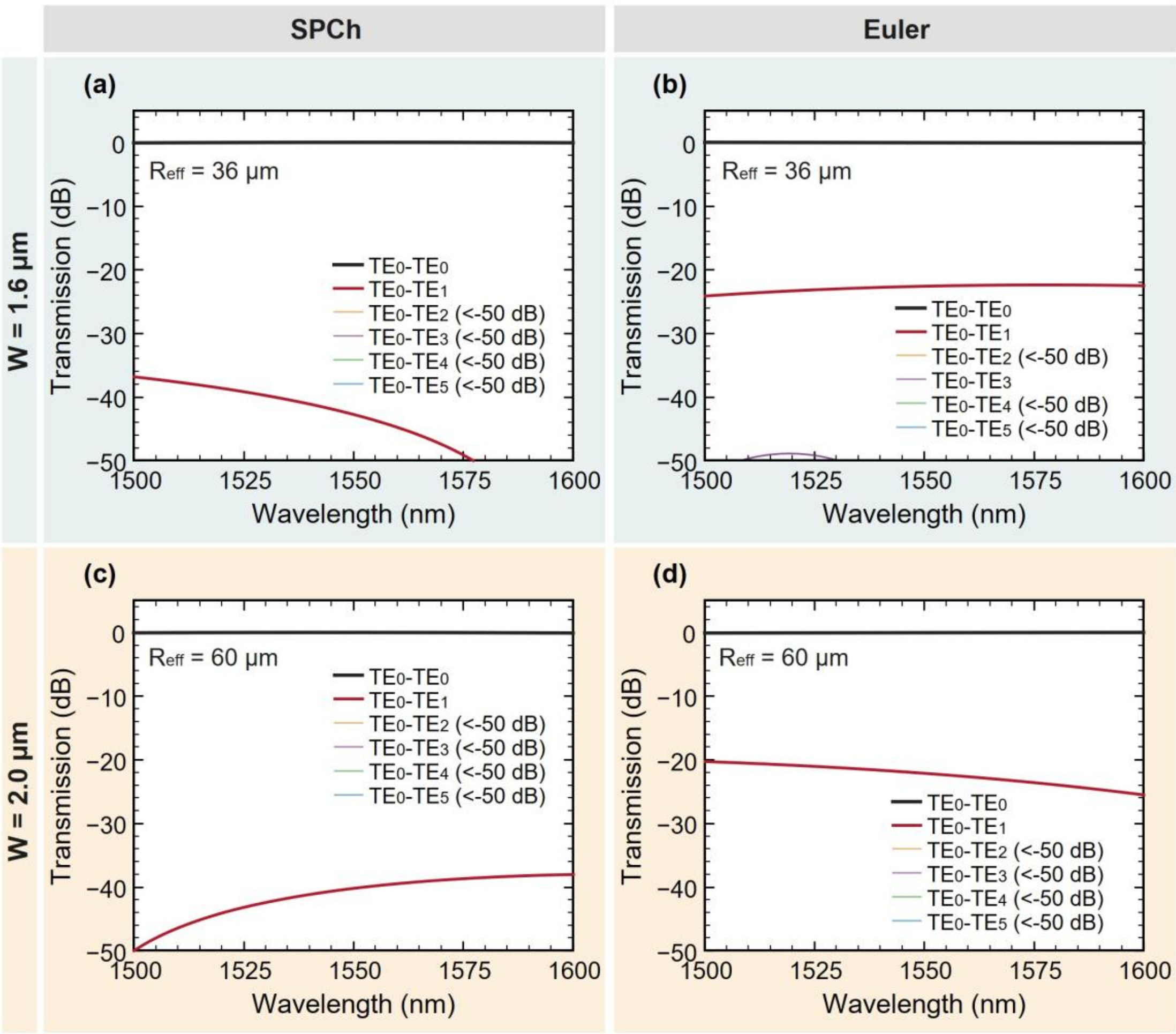

**Figure S3.** Wideband response of relaxed 90° bends. Transmission spectra of 90° bends with $TE_0$ launched and the output decomposed onto $TE_0$–$TE_5$ across 1500–1600 nm. **a**, SPCh bend at $W$ = 1.6 μm and $R_{\mathrm{eff}}$ = 36 μm. **b**, Euler bend at $W$ = 1.6 μm and $R_{\mathrm{eff}}$ = 36 μm. **c**, SPCh bend at $W$ = 2.0 μm and $R_{\mathrm{eff}}$ = 60 μm. **d**, Euler bend at $W$ = 2.0 μm and $R_{\mathrm{eff}}$ = 60 μm.

---

## S6. Tunable single-passband microwave photonic filter

The high quality factor ($Q$) enabled by the SPCh bends translates to narrow-band microwave photonic filtering. The $W$ = 2.0 μm SPCh-MRR (Device C in Table 1 of the main text) is integrated into a tunable single-passband microwave photonic filter (MPF) based on phase-modulation-to-intensity-modulation (PM–IM) conversion [S2, S3]. The MRR itself acts as an optical notch resonator. In the PM–IM scheme, this notch disrupts inter-sideband cancellation of the phase-modulated optical signal. The resulting amplitude imbalance is detected by a photodetector to form a radio-frequency (RF) passband centered at frequency $F_c$. An integrated thermo-optic heater tunes the MRR resonance and shifts $F_c$ accordingly.

Fig. S4a shows the normalized RF transmission (the forward scattering parameter $S_{21}$) of the resulting MPF, with passbands tuned across 2.26–42.79 GHz, plotted on a viridis colormap parameterized by $F_c$. Fig. S4b zooms in on a representative passband, where a Lorentzian fit yields a 3-dB bandwidth of 76 MHz. Together, the 40-GHz tuning range and the narrow passband confirm that the SPCh-MRR serves as a high-$Q$ building block for compact, programmable RF-photonic processing.

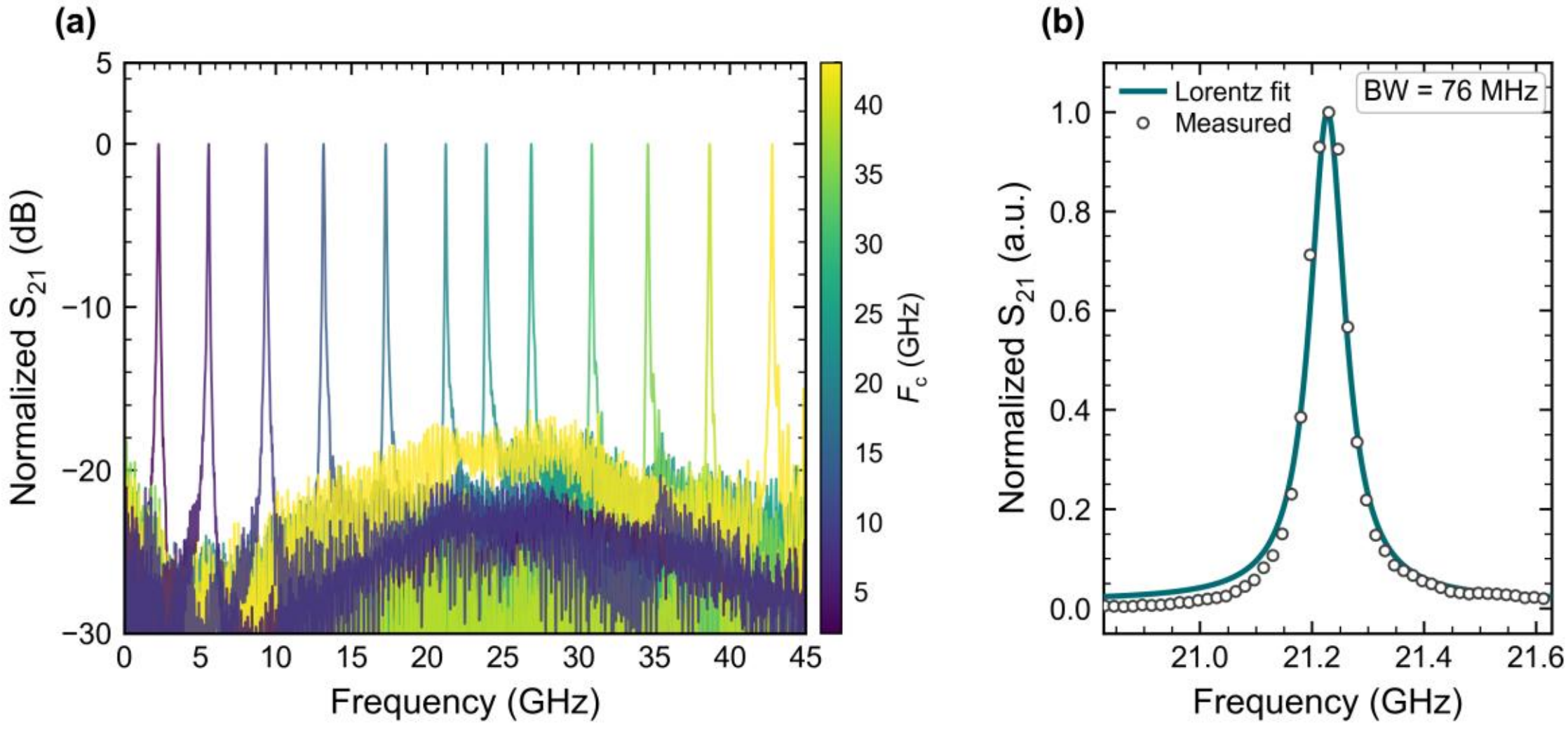

**Figure S4.** Tunable single-passband MPF based on the SPCh-MRR. **a**, Normalized $S_{21}$ of single-passband responses with center frequency $F_c$ across 2–43 GHz. **b**, Zoomed-in view of a representative passband at $F_c \approx$ 21 GHz (±400 MHz span). A Lorentzian fit to the linear-amplitude $S_{21}$ yields a 3-dB bandwidth of 76 MHz.

---

# S7. Cross-platform validation on x-cut LNOI

The infinite-differentiability necessary condition established in the main text depends only on the polynomial structure of the Lagrangian density. The SPC framework should therefore transfer to other multimode platforms. To verify this, we performed mode-decomposition simulations on x-cut LNOI [S4, S5]. LNOI supports high-$Q$ microring resonators on integrated chips [S6]. The effective bend radius was fixed at 100 µm, and two typical multimode waveguide widths on LNOI were chosen: $W$ = 1.5 µm and $W$ = 2.5 µm. The waveguide consists of a 600-nm-thick LN ridge on a buried oxide, partially etched 300 nm into the film and leaving a 300-nm slab, with air as the top cladding. The ridge sidewalls are inclined at 60°, matching the typical angle obtained by inductively coupled plasma etching of x-cut thin-film lithium niobate. Because the lower index contrast of LNOI (Δn ≈ 0.7) requires a larger $R_{min}$ than SOI to confine the $TE_0$ mode, we used $R_{min}$ = 50 µm at both widths.

Fig. S5 presents the transmission spectra of Euler and SPCh bends across 1500–1600 nm, including the $TE_0$ transmission along with the lowest higher-order $TE_1$ mode and the transverse-magnetic (TM) modes $TM_0$ and $TM_1$.

SPCh maintains $MER_{max}$ at −36.3 dB ($W$ = 1.5 μm) and −37.5 dB ($W$ = 2.5 μm), with $IL_{max}$ below 0.006 dB at both widths. At $W$ = 1.5 μm, the Euler bend reaches $MER_{max}$ = −29.7 dB at the same $R_{min}$, corresponding to 6.6 dB lower mode suppression than SPCh. At $W$ = 2.5 μm, the gap widens: the Euler bend reaches only $MER_{max}$ = −27.8 dB, a 9.7 dB deficit. A wider waveguide supports more guided modes (six at $W$ = 2.5 μm versus four at $W$ = 1.5 μm) under the present ridge geometry. Adiabatic transmission therefore becomes more demanding as the core widens, and the SPCh advantage over the Euler bend grows with width, mirroring the trend observed on SOI. $TE_1$ remains the dominant leakage channel in three of the four panels, whereas the wider 2.5 μm SPCh design is limited by the polarization-coupling channel $TM_0$.

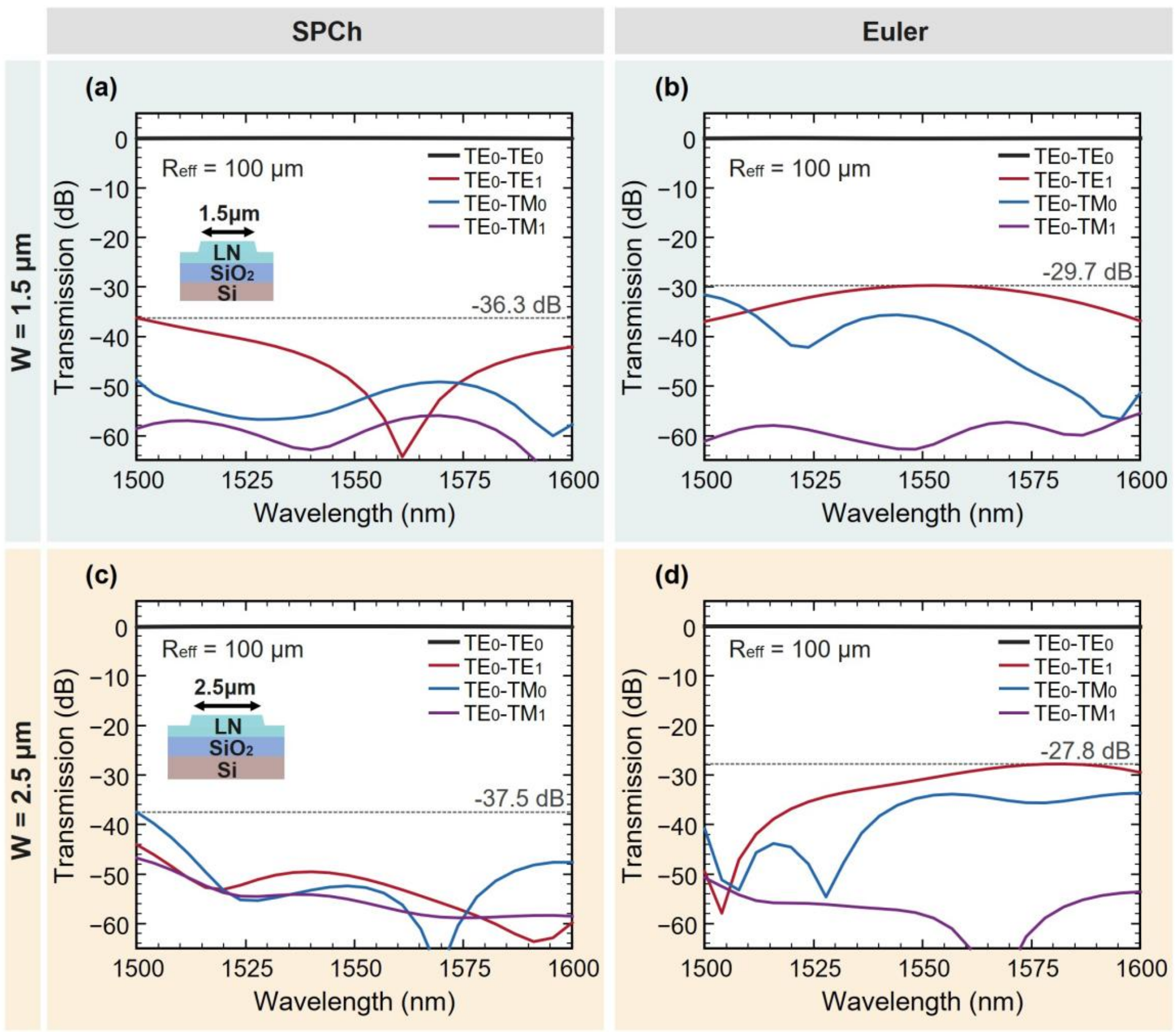

**Figure S5.** Platform generality on x-cut LNOI. Transmission spectra of 180° bends on x-cut LNOI at $R_{eff}$ = 100 μm, with $TE_0$ launched and the output decomposed onto $TE_0$, $TE_1$, $TM_0$, and $TM_1$ across 1500–1600 nm. **a**,

SPCh bend at $W$ = 1.5 μm. **b**, Euler bend at $W$ = 1.5 μm. **c**, SPCh bend at $W$ = 2.5 μm. **d**, Euler bend at $W$ = 2.5 μm. The dashed gray lines mark the dominant higher-order excitation.

---

## S8. Geometric construction of SPC bends

An SPC bend is assembled directly from its designed curvature profile $\kappa(s)$, where $s$ is the arc length measured along the bend (Fig. S6a). Integrating the curvature yields the cumulative tangent angle $\varphi(s) = \int_0^s \kappa(t)\,dt$, which increases monotonically along the bend (Fig. S6b). For a target bend angle $\theta$, the half-bend arc length $s_0$ is fixed by requiring the tangent to turn through $\theta/2$ over the half-bend,

$$\varphi(s_0) = \int_0^{s_0} \kappa(s)\,ds = \theta/2. \tag{S24}$$

Fig. S6 uses $\theta$ = 180° as an example, and other bend angles follow the same procedure with $\theta$ set to the desired value. The centerline coordinates then follow by trapezoidal integration of the tangent direction,

$$x(s) = \int_0^s \cos\varphi(t)\,dt, \tag{S25a}$$

$$y(s) = \int_0^s \sin\varphi(t)\,dt, \tag{S25b}$$

where $\varphi$ is the tangent angle defined above. The full bend is symmetric about its midpoint, so its curvature satisfies $\kappa(2s_0 - s) = \kappa(s)$, with $s_0$ the half-bend arc length and $2s_0$ the full arc length. The second half therefore runs from $\kappa_{\max}$ at the midpoint back to zero at the output, keeping $\kappa$ continuous across the midpoint and zero at both straight-to-bend junctions. Offsetting the centerline by $\pm W/2$ at half the design width gives the two waveguide boundaries, producing the closed polygon of Fig. S6c that is compatible with both Graphic Design System II (GDSII) export for fabrication and direct import into finite-difference time-domain (FDTD) models. Because every step integrates the smooth $\kappa(s)$, the resulting outline inherits its continuity, so an SPC bend is free of the curvature breaks that a circular arc or an Euler bend introduces at the ports.

A valid S-bend consists of two full bends, each running from $\kappa$ = 0 to $\kappa_{\max}$ and back to 0, joined in a point-symmetric configuration with $\kappa$ = 0 at the junction. This construction preserves full curvature continuity.

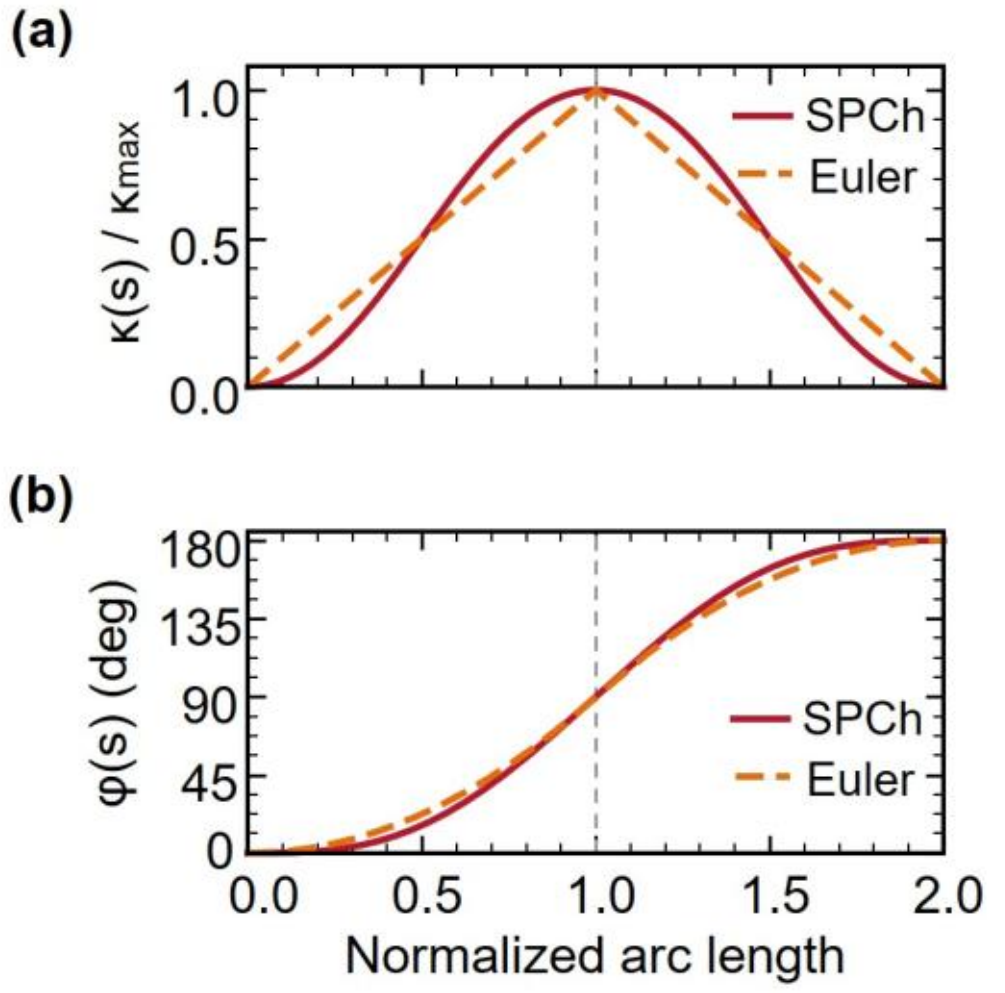

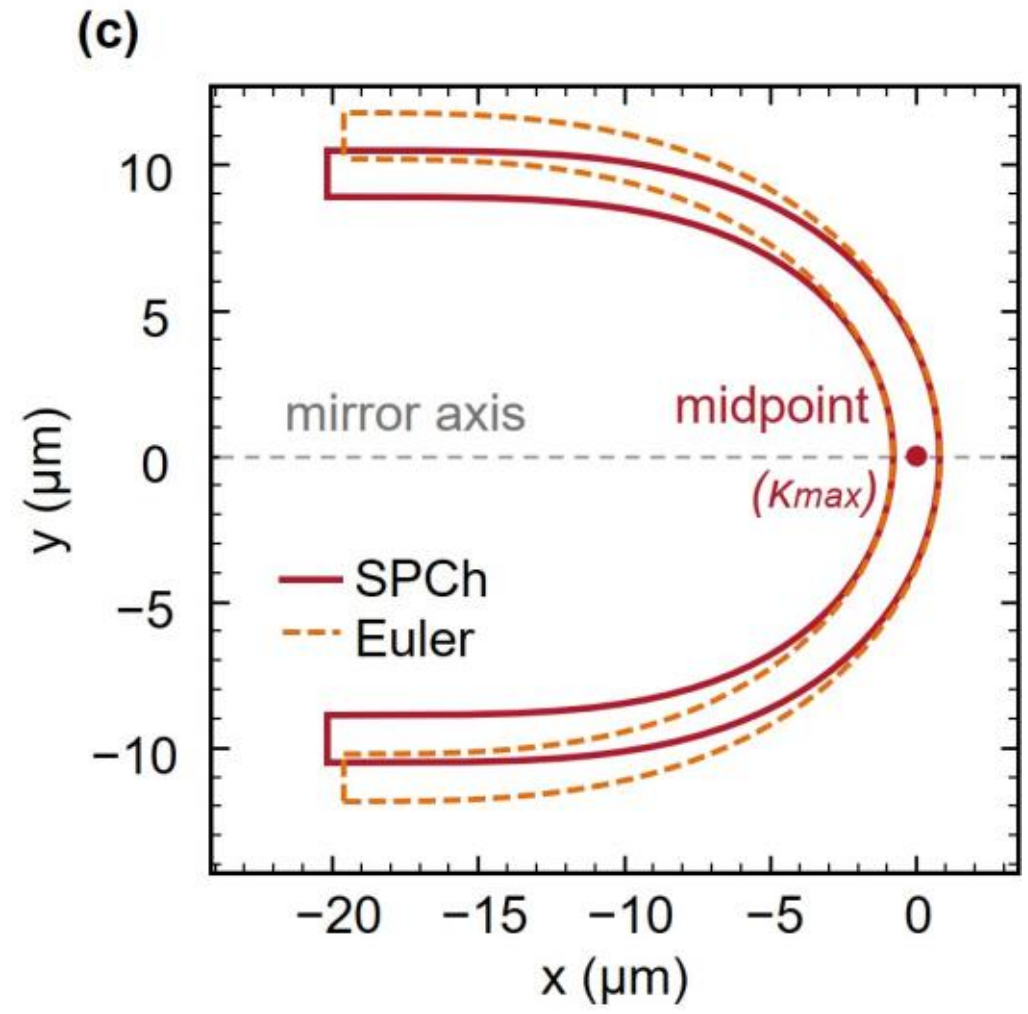

**Figure S6.** Geometric construction of an SPC bend. **a**, Designed curvature profile $\kappa(s)/\kappa_{\max}$ along the normalized arc length for the SPCh and Euler bends, mirror-symmetric about the midpoint (dashed line). **b**, Cumulative tangent angle $\varphi(s) = \int_0^s \kappa \, dt$, rising from 0° to 180° across the full bend. **c**, The resulting 180° waveguide outline, obtained by integrating the centerline $(x, y) = \int (\cos\varphi, \sin\varphi)\, ds$ and offsetting it by $\pm W/2$ ($R_{\min}$ = 8 μm, $W$ = 1.6 μm). The SPCh (solid) and Euler (dashed) outlines are aligned at the midpoint of maximum curvature, about which each bend is mirror-symmetric (dashed horizontal axis).

---

## Supporting References

[S1] Melati, D., Morichetti, F. & Melloni, A. A unified approach for radiative losses and backscattering in optical waveguides. *J. Opt.* **16**, 055502 (2014).

[S2] Garrett, M. et al. Integrated microwave photonic notch filter using a heterogeneously integrated Brillouin and active-silicon photonic circuit. *Nat. Commun.* **14**, 7544 (2023).

[S3] Feng, H. et al. Integrated lithium niobate microwave photonic processing engine. *Nature* **627**, 80–87 (2024).

[S4] Wang, C. et al. Integrated lithium niobate electro-optic modulators operating at CMOS-compatible voltages. *Nature* **562**, 101–104 (2018).

[S5] Hu, Y. et al. Integrated electro-optics on thin-film lithium niobate. *Nat. Rev. Phys.* **7**, 237–254 (2025).

[S6] Zhu, X. et al. Twenty-nine million intrinsic Q-factor monolithic microresonators on thin-film lithium niobate. *Photonics Res.* **12**, A63–A68 (2024).